\documentclass{aa}  

\usepackage{graphicx}
\usepackage{txfonts}
\begin{document}

   \title{Migration of Jupiter mass planets in low viscosity discs}

   \subtitle{}

   \author{Lega, E.\inst{1}, Nelson R.P.\inst{2}, Morbidelli, A.\inst{1}, Kley, W. \inst{3}, B\'ethune, W. \inst{3}, Crida, A.\inst{1}, Kloster, D.\inst{1}, M\'eheut, H.\inst{1}, Rometsch, T.\inst{3}, Ziampras, A.\inst{3}
          }

   \institute{Laboratoire Lagrange, UMR7293, Université de Nice Sophia-Antipolis, CNRS, Observatoire de la Côte d'Azur, Boulevard de l'Observatoire, 06304 Nice Cedex 4, France. \and Astronomy Unit, School of Physics and Astronomy, Queen Mary University of London, London E1 4NS, UK \and Institut für Astronomie und Astrophysik, Universität Tübingen, Auf der Morgenstelle 10, 72076, Tübingen, Germany}

   \date{Received S; accepted }

% \abstract{}{}{}{}{} 
% 5 {} token are mandatory
 
  \abstract
 {Type-II migration of giant planets has a speed proportional to the disc's viscosity for values of the $\alpha$ viscosity parameter larger than $10^{-4}$ \citep{Robertetal18}. At even lower viscosities  previous studies,  based on 2D simulations \citep{LinPapaloizou10,LinPapaloizou11,2017ApJ...839..100F,McNallyetal19}, have shown that migration can be very chaotic and often characterized by phases of fast migration. The reason is that in low-viscosity discs vortices appear due to the Rossby-wave instability at the edges of the gap opened by the planet. Migration is then determined by vortex-planet interactions.  
}
{Our goal is to study giant planet migration in low-viscosity discs with 3D simulations. In 3D,  vortices are more complex than the simple vertical extension of their 2D counterparts \citep {Barranco and Marcus 2005, Meheut et al. 2010}; their impact on planet's migration is therefore not obvious.}
{We performed  numerical simulations using 2 grid-based codes: FARGOCA for 3D simulations and  FARGO-ADSG for the 2 dimensional case.  2D-simulations were used mainly for preliminary tests to check the impact of self-gravity on vortex formation and on vortex-disc dynamics. After selecting disc masses for which self-gravity is not important at the planet location, 3D simulations without self-gravity can be safely used. We have considered an adiabatic equation of state with exponential damping of temperature perturbations in order to avoid the development of the vertical shear instability.  In our nominal simulation we set  $\alpha = 0$ so that only numerical viscosity is present. We then performed simulations with non zero 
$\alpha$ values to assess the threshold of prescribed viscosity below which the new migration processes appear.}
{We show that for $\alpha \lesssim 10^{-5}$ two migration modes are possible which differ from classical Type-II migration, in the sense that they are not proportional to the disc's viscosity. The first occurs when the gap opened by the planet is not very deep. This occurs in 3D simulations and/or when a big vortex forms at the outer edge of the planetary gap, diffusing material into the gap. The desaturation of coorbital and corotation resonances \citep{GoldTre1980,GoldSari03} keeps the planet's eccentricity low. Inward planet migration then occurs as long as the disc can refill the gap left-behind by the migrating planet, either due to diffusion caused by the presence of the vortex or to the inward migration of the vortex itself due to its interaction with the disc. We call this type of migration "vortex-driven migration", which differs from "vortex-induced" migration described in \cite{LinPapaloizou10,LinPapaloizou11}.   This migration is very slow and cannot continue indefinitely, because eventually the vortex dissolves. The second migration mode occurs when the gap is deep so that the planet's eccentricity grows to a value $e \sim 0.2$ due to inefficient eccentricity damping by corotation resonances. Once the planet is on an eccentric orbit, gas can pass through the gap and planet migration unlocks from the disc's viscous evolution  \citep{Duffelletal14,Durmann and Kley 2015}. This second, faster migration mode appears to be typical of 2D models in discs with slower damping of temperature's perturbations.
 }
   {Vortex-driven migration in low-viscosity discs can be very slow and eventually reverses and stops, offering an interesting mechanism to explain the existence of the cold-Jupiter population, even if these planets originally started growing at the disc's snowline.}

   \titlerunning{Giant planet migration in inviscid discs}
   \authorrunning{E. Lega et al.}

   \maketitle
%
%-------------------------------------------------------------------

\section{Introduction}
\label{sec:Intro}
Understanding the origins of giant planets remains elusive. Radial velocity surveys
have found giant planets to exist around roughly 10\% of Sun-like stars \citep{Mayor2011, Cumming2008}. However, only $\sim 1$\% of Sun-like stars have hot Jupiters on very short-period
orbits \citep{Howard2010Sci}.  Very few stars have warm Jupiters with orbital radii of up
to 0.5-1 au \citep{Butler2006, Udry2007}. Instead, { when considering the unbiased distribution, } most giant planets are found  between 1 and several au { \citep{Butler2006, Udry2007,Cumming2008,Howard2010Sci,Mayor2011}}, while there are hints that their number decreases again farther
out \citep{Mayor2011,Fernandes2019}. In our Solar System, of course, there are no giant planets within 5 AU from the Sun, although Jupiter may have been at $\sim 2$~au in the past \citep{GT2011}. The preference for giant planets to orbit relatively far from the parent star, in contrast to super-Earths for example, is puzzling because giant planets must have formed in the presence of gas in the protoplanetary disc and, consequently, they should have migrated towards the central star due to planet-disc interactions. 

Giant planet migration, a.k.a. Type-II migration, has been the subject of a rich literature since the pioneering work by \citet{LinPap1986}. In essence, a giant planet opens a gap in the gas around its orbit. Any migration of the planet has to occur in concert with readjustment of the gas in the disc, so that the gap can migrate together with the planet. In a viscous accretion disc, this can happen only on a viscous timescale. Hence Type-II migration is expected to be inwards, with a radial speed of order $v_r \sim 1.5\nu/r$, where $\nu$ is the viscosity, i.e. with the same radial speed at which the gas viscously accretes towards the central star \citep{Ward1997}. Young stars typically accrete gas at a rate of $\sim 10^{-8}M_\odot$/y (with a large, order of magnitude scatter around this value \citep{1998ApJ...495..385H,Manara2016}; for a density of gas in the disc at 1~au comparable to that of the Minimum Mass Solar Nebula model ($2\times 10^{-4}M_\odot$/au$^2$; \citet{Stu1977,Hayashi1981})  this implies a radial speed of $\sim 8$~au/My. This means that, at Type-II migration speeds in viscous discs, giant planets should move towards their host stars on timescales much shorter than disc's lifetimes. If this were true, the presence of giant planets at several au from the parent star would require that these planets formed very far away (i.e. beyond $\sim 20$~au) and/or quite late in the history of the protoplanetary disc \citep{ColemanNelson2014,Bitschetal2015}. 

However, this simple solution is not without problems. The sweetspot for the rapid formation of massive cores capable of accreting gas and becoming giant planets is the snowline \citep{2017A&A...602A..21S, 2017A&A...608A..92D}. If the giant planets we observe today orbiting beyond 1~au formed at distances $\ge 20$~au, where are the planets that should have formed faster and in larger numbers near the snowline? A popular idea was that these planets have fallen onto the central star and the planets that we observe now are the last of the Mohicans \citep{Lin1997,LaugAdam1997}. Today this idea has been mostly discarded because of the pile-up of extrasolar planets on orbits with periods of 3--10 days, and modern models of interactions between stellar magnetospheres and protoplanetary discs \citep{2005ApJ...632L.135M,2008ApJ...687.1323M,2012ApJ...744...55A} suggest that discs are truncated at a few $10^{-2}$~au, which should prevent planet migration continuing all the way to the stellar surface. 

If the giant planets we observe formed at the snowline, they must have migrated only a few au during the lifetime of the disc. Because the rate of Type-II migration is in principle proportional to the disc's viscosity, this may suggest very low viscosities in protoplanetary discs. This idea is supported by modern studies on turbulent viscosity in discs. Turbulence was originally expected to arise from the magneto-rotational instability (MRI; \citet{Balbus1991}). However, it was later understood that the ionization of the gas near the midplane of the disc is too weak to sustain the MRI \citep{1996ApJ...457..355G,Stone1996}, introducing the concept of the dead zone. Even more recently, the inclusion of non-ideal MHD effects, such as ambipolar diffusion, led to the conclusion that the coupling between the magnetic field and the gas should not make the disc turbulent even at its surface (see \citet{Turner2014} for a review). Another often considered source of turbulence, the vertical shear instability (VSI hereafter; \citet{2013MNRAS.435.2610N,2014A&A...572A..77S}) should also not be active at the disc midplane within a few au from the star because the disc's cooling rates are too slow (however, see \citet{2020arXiv200811195P} for a different view). Thus, the idea of formation and migration of giant planets in low viscosity discs may be appealing.

Unfortunately, the dependence of Type-II migration on viscosity in the limit of small viscosity is far from clear. \citet{Duffelletal14} claimed that the paradigm of Type-II migration is flawed because the gas can pass through the planet's orbit, from one side of the gap to the other, so that the planet is not locked in the viscous evolution of the disc. For this reason, the planet's migration speed can be different from  $-1.5 \nu/r$ where $\nu$ is the viscosity. \citet{Durmann and Kley 2015} confirmed this result, but \citet{Robertetal18} showed that the passage of gas through the gap is inhibited at small viscosity because the gap is much larger than the planet's horseshoe region; consequently the planet's migration rate remains proportional to $\nu$ although not necessarily equal to $-1.5 \nu/r$. However, all these studies have been performed using 2D simulations and, more importantly, for numerical reasons considered viscosities that were not very small. Adopting the usual prescription $\nu=\alpha H^2\Omega$ \citep{SS73}, where $H$ is the pressure scale-height of the disc and $\Omega$ is the orbital frequency, the viscosities considered in these works corresponded to $\alpha > 10^{-4}$. For these values of $\alpha$, Type-II migration, even if proportional to $\nu$, is still too fast to explain the current location of most giant planets, if these planets formed at the snowline. At face value, $\alpha$ should be $\sim 10^{-5}$ for Type-II migration to be slow enough, if the proportionality between migration speed and $\alpha$ (or $\nu$) holds also for such a low viscosity. 

Attempts to simulate planet migration in inviscid discs have led to very chaotic and erratic planetary evolutions, often characterized by phases of  very fast migration \citep{LinPapaloizou10,LinPapaloizou11, McNallyetal19}. The reason is that in low viscosity discs with embedded giant planets large scale vortices can appear due to edge instabilities of planetary gaps and Rossby wave instabilities at pressure bumps \citep{Lovelaceetal99,Kolleretal03,LinPapaloizou10}. The interaction of the planet with a vortex can lead to several effects. For instance, the passage of the vortex from one side of the gap to the other can accelerate the planet's migration in the opposite direction (called ''vortex-induced'' migration in \citet{LinPapaloizou10}). A further complication is that gap-opening planets in inviscid discs may undergo eccentricity growth because of (i) their interaction with the vortices and (ii) the saturation of corotation resonances that are normally responsible for damping eccentricity \citep{GoldTre1980,GoldSari03}; in turn the planet's eccentricity can have a strong feedback on  planet migration.

However, it has been shown that the formation and evolution of vortices depends sensitively on the simulation set-up. In 2D simulations,  \citet{ZhuBaruteau16} showed that vortices generated by the Rossby wave instability are less pronounced if the self-gravity of the disc is taken into account.  Using 2D models \citet{LinPapaloizou11,LinPapaloizou11b} showed that, when considering self-gravity in discs with suitable mass, the vortex-induced migration observed in \citet{LinPapaloizou10} is considerably delayed. In 3D models, vortices are more complex than the vertical extension of their 2D counterparts \citep{Barranco and Marcus 2005,Meheut et al. 2010}. Thus the interplay between planets and vortices and the effects of this interplay on migration are far from clear. 

In this context, the goal of this paper is to provide a comprehensive exploration of giant planet migration in  low-viscosity discs. We approach this problem with a suite of 2D simulations with and without
 disc self-gravity and 3D simulations without self-gravity. More specifically, after a brief presentation of our physical models (Sect. 2) and of the simulation set-ups (Sect.~3), we present in section 4 the structure of an inviscid disc under the effect of a Jupiter-mass planet kept on a fixed circular orbit at 5.2 au, with emphasis on the generation of vortices and their evolution. Then, in section 5 we investigate the planet's migration in inviscid discs using both 2D and 3D simulations. We also provide the range of validity of our results by finding the transition to  classical type II migration. We show in section 6 that results depend on a planet's formation site. Numerical tests such as: convergence with respect to resolution and gravitational smoothing of the potential in the planet vicinity as well as a discussion of dependence of results on the choice of the equation of state are discussed in the Appendix.
The conclusions and a global discussion of the results are reported in Sect.~7.

\par In a subsequent paper, we will continue investigating giant planet migration in inviscid 3D discs in two directions:
i) taking into account self-gravity in 3D discs for the study of migration of giant planets forming at large distances from the star ($\sim 20$~au)
ii)  modeling the effect of the radial advection of gas in the disc
 due to angular momentum removal by magnetised disc winds  \citep{BaiStone2013,Turner2014,2015ApJ...801...84G,Bethuneetal17}.
 Advection is required in a realistic model of low viscosity discs to provide a mass flux to the star comparable to the observed stellar accretion rates.
 The effect of advection on giant planets migration will be the object of a forthcoming paper.
Nevertheless, a deep understanding of planet migration in low-viscosity discs with no advection, as that developed in this paper, is  required in order to understand giant planet migration in a more realistic scenario.

%--------------------------------------------------------------------

\section{Physical models} 
We briefly describe our 2D and 3D models in this section. 

\subsection{The 3D model}
The protoplanetary disc is treated as a non self-gravitating gas whose motion is described by the Navier-Stokes equations.
We use spherical coordinates $(r,\varphi,\theta)$ 
where $r$ is the radial distance from the star (which is at the origin of the coordinate system), $\varphi$ is the azimuthal coordinate measured from the $x$-axis and $\theta$ the 
polar angle measured from the $z$-axis (the colatitude). The midplane of the disc is located at the equator
$\theta = \frac {\pi} {2}$. 
We work in a coordinate system which rotates with angular velocity:
$$\Omega_0 = \sqrt {\frac{GM_{\star}}{{r_p(0)}^3}} $$
where $M_{\star}$ is the mass of the central star, $G$ is the gravitational constant,
and $r_p(0)$ is the initial distance to the star from a planet of mass $m_p$, assumed to be on a circular orbit.
 The gravitational influence of the planet on
the disk is modelled as in \citet{KBK09} using the full gravitational
potential for disc elements having distance $d$ from the planet larger than a fraction $\epsilon$ of the Hill radius,
and a smoothed potential for disc elements with $d<\epsilon$.

\par
We integrate the Navier-Stokes equations taking into account indirect forces that account for the acceleration of the star by the disc and planet \citep{Masset2002}. As shown in \citet{ZhuBaruteau16}, indirect forces have an impact on shaping vortices that is proportional to the disc mass.
We add an equation for the internal energy $e=\rho c_v T$ to the Navier-Stokes equations, where $\rho$
and $T$ are the volume density and the temperature of the disc gas and
$c_v$ is the specific heat at constant volume:
\begin{equation}
\label{energyeq}
\frac{\partial e} {\partial t} + \nabla \cdot (e\vec v)
 =  -p\nabla \cdot \vec v  - c_v\rho \frac{T-T_0}{\tau_c},
\end{equation}
where  $\tau_c$ is the cooling time and $T_0$ is the initial
temperature, defined as $T_0(r)=GM_*\mu h_0^2/(R_{gas}r)$, with $h_0$ being the disc
aspect ratio, $\mu$ the mean molecular weight ($\mu=2.3 {\rm g}/{\rm mol} $ for a standard solar mixture) and $R_{gas}$ is the ideal-gas constant. \par In short, we use
an adiabatic EoS, on top of which we exert an exponential damping  of the temperature
perturbations. We do not use the simpler locally isothermal EoS in order to avoid disc instabilities
like the VSI.
In the quoted paper it is shown that a disc with an equation of state taking
into account  thermal relaxation like in Eq.\ref{energyeq} with suitable values of $\tau _c$ is not
prone to develop such an instability.
 According to the same paper we will consider $\tau_c$ equal to 1 orbital period at the planet location. A self-consistent simulation of the disc's evolution indeed shows that no significant VSI should  develop at Jupiter's distance (Ziampras et al., in preparation, but see  \citet{2020arXiv200811195P} for results with a different simulation set-up.). The results for different cooling times are reported in Appendix \ref{sec:Appendix3D}.
\subsection{The 2D model}
Two-dimensional models have been commonly used in the literature for studies of planetary migration. Evolution over long timescales (up to hundreds of thousands of planetary orbits) is possible at reasonable computational cost using 2D simulations, which makes them useful for exploring long term behaviour. The obvious disadvantage is that possible genuine 3D effects cannot be observed.
\par
%Actually, protoplanetary discs are vertically thin so they can be
%approximated  as 2 dimensional annulii described in polar coordinates $(r,\varphi)$ and in many cases three dimensional effects have negligible impact on planet-disc interactions.\par
In the two-dimensional model we solve the vertically-integrated Navier-Stokes equations using polar coordinates $(r,\varphi)$ and, similar to the 3D model described above, we also solve equation~(\ref{energyeq}) with thermal relaxation, where in 2D the internal energy density is defined by $e=\Sigma c_v T$, and $\Sigma$ is the surface density.
We also consider a cooling time equal to 1 orbital period at the planet position, and we will discuss in Appendix \ref{sec:Appendix2D} the sensitivity of results to the choice of the cooling time.
\par
 We undertake 2D simulations that either include or neglect the effects of the disc self-gravity. Vortices are an inevitable consequence of gap opening by planets in inviscid discs, and their evolution is expected to depend on the relative influences of the indirect term and self-gravity \citep{ZhuBaruteau16}. Hence, it is important to test the effects of self-gravity and to determine under which conditions it is important or can be neglected. 
\par
The 2D simulations are performed in a frame of reference in which the star is located at the origin, and which rotates at a rate that corresponds to the instantaneous angular velocity of the planet (i.e. corotating with the planet even when the planet migrates).
 
\section{Setup of numerical simulations}
 Our 3D simulations are done with the code FARGOCA ( FARGO with {\bf C}olatitude {\bf A}dded;
 \citet{Lega14})\footnote{The simulations presented in this paper have been obtained with a recently re-factorized version of the code that can be found at: https://disc.pages.oca.eu/fargOCA/public/}. The code is based on  the  FARGO code \citep{Masset00} extended  to 3 dimensions. The fluid equations are solved using  a second order upwind scheme with a
time-explicit-implicit multistep procedure.
The code is parallelized using a hybrid combination of MPI between the
nodes and OpenMP on shared memory multi-core processors. 
The two dimensional simulations are done with FARGO-ADSG, a version of the FARGO code which implements an energy equation and disc self-gravity as explained in \citet{BarMas08}.
 
The code units are $G=M_*=1$, and the unit of distance $r_1=1$ is arbitrary when expressed in au.
The unit of time is therefore $r_1({\rm au})^{3/2}/(2\pi)\,{\rm yr}$. 
For the simulations in this paper we adopt the Sun-Jupiter distance as the unit of length in au: $r_1=5.2$. When presenting simulation results distances are expressed in au and time in years.

\subsection{Disc parameters}
\label{sec:3.1}
Unless otherwise stated, all models have a radial domain defined by $r_{\rm min}\leq r \leq r_{\rm max}$ with $r_{\rm min}=1.04$ au and $r_{\rm max}=46.8$ au. Such a wide radial extension is quite unusual in these kind of studies. The paper focuses on the case of a planet initially at 5.2~au but, in order to test the dependence of the results on the planet-star distance,  we also considered the case of
a planet initially at 13~au (see section \ref{sec:FarPlanets}), and in this case we found that an extended radial domain is required to ensure that results are not affected by the outer boundary. In the radial direction we use the classical prescription \citep{ValBorro2006}) of evanescent boundary condition.

Importantly, in our nominal simulations, we set $\alpha=0$, i.e. we consider inviscid discs. Of course, there will still be some resolution-dependent numerical diffusivity in the code. This issue will be discussed  in subsection 5.4 , but we anticipate here that, for our nominal resolution, the numerical viscosity will not be larger than the one in a viscous disc with $\alpha=10^{-5}$. Moreover, in all the simulations
(2D and 3D) we consider artificial viscosity to stabilize shocks (precisely von Neumann-Richtmyer viscosity and corresponding heating terms
 as described by \citet{StoneNorman1992}).

In the 3D models, the meridional domain extends from the midplane to $12^{\circ}$ above the midplane ($\theta=78^{\circ}$--$90^\circ$), about 4 disc scale-heights. We do not study inclined planets and therefore we do not need to extend the domain below the midplane. Mirror boundary conditions are applied at the midplane as in \citet{KBK09} and reflecting boundaries are applied at the disc surface.    The initial temperature profile is the same for all the 3D discs: isothermal in the vertical direction, while in the radial direction the temperature scales as $1/r$. This gives rise to a model in which the disc aspect ratio, $h$, is constant with radius, and we adopt the value $h=h_0=0.05$.
\par
 
 In the 2D models we adopt a radial profile for the disc aspect ratio given by \citet{Chiang and Goldreich (1997)} for a passively heated disc: $h=h_0(r/r_1)^{2/7}$ with $h_0=0.05$.  This choice was made because in earlier test calculations undertaken with $h=0.05$ throughout the disc, the value of the Toomre $Q$ parameter approaches unity at the edge of the disc models, such that the effects of self-gravity on global disc evolution became very obvious in the form of global spiral waves being excited. Adopting a variable $h$ removed this problem.\par
 {We remark that the aspect ratio is the same in 2D and 3D models at the planet's location, 5.2~au, and is slowly diverging away from this radius. The aspect ratio does not change during the simulations.\footnote{The nominal cooling time  $\tau_{\rm c}=1$~orbital period effectively damps temperature perturbations} This is true also for simulations with non zero viscosity (see Section 5.4). Precisely, we do not include the term corresponding to  the viscous heating in Eq.\ref{energyeq} in order to have the same vertical structure for all the simulations and make them comparable.}
 \par
The disc surface density in 2D and 3D models is given by
$\Sigma=\Sigma_0(r/r_1)^{-1/2}$, with $\Sigma_0= 6.76\times10^{-4}$ in code units for our nominal disc (corresponding to $222\, \mbox{g\, cm}^{-2}$ at 5.2~au). We will call the corresponding simulations ``nominal'' or $M$ simulations (see table\ref{table:1}).
Classic Type-II migration is not dependent on the disc mass, except in the inertial limit where the local disc mass is much smaller than the planet mass and migration slows down \citep{Quillen04}. However, in a low viscosity disc, in which vortices can form at the edge of the planet-induced gap, the mass of the disc may play an important role. In particular, \citet{ZhuBaruteau16} showed that the strength and evolution of vortices depends on both the gravitational acceleration that they exert on the central star (as expressed through the indirect term) and the self-gravity of the disc.
Therefore, we also run  simulations with the nominal disc mass divided by some integer $n$, we will call these simulations $M/n$ 
or $\rm Nominal/n$.

Achieving long run times for 3D simulations is challenging in terms of computational cost, therefore we have used a moderate resolution of $(N_r,N_{\theta},N_{\varphi})=(568,16,360)$
with uniform radial grid spacing. 
{At this resolution we  consider a smoothing length $\epsilon = 0.8 R_H$,
i.e.  4 grid-cells in the smoothing length. Although this value is quite large, it has been shown  in  \cite{Fungetal2017} that convergence in the torque measurements in non viscous disks requires at least 3 grid cells per smoothing length. We provide a test in the Appendix with  smoothing length $\epsilon = 0.4 R_H$
at the same resolution and show that results are converged, and we further reduce the smoothing length ($\epsilon = 0.2 R_H$)  in the  case where we double the resolution (see Appendix \ref{resolution}).}
We  note that in our comparison between 2D self-gravitating and non-self-gravitating discs, we adopt the same number of cells in the radial and azimuthal directions as in the 3D runs, but the radial grid spacing is logarithmic, as is required by the self-gravity solver in FARGO-ADSG.
Table 1 lists the principal parameters and simulation names to which we will refer in the following.{ We notice that viscosity is labeled with the
$\alpha$ parameter. Since the vertical structure remains constant in the simulations $\alpha$ gives directly the kinematic viscosity $\nu$ through the relation $\nu = \alpha h^2 \Omega$.}
\begin{table}
\caption{Simulation  parameters for 3D and 2D discs: the simulation names capture the main parameters to help the readability of results. In the 2 dimensional cases the index sg is added to simulations taking into account self-gravity. The index $13$~au indicates the case of a distant planets, for all the others simulations the planet is initially set at $5.2$~au. We have also run control simulations for smaller-mass discs, we do not list them  in the table,  we will call these simulations $M/n$ 
or $\rm Nominal/n$ with $n=2,4,10$}             % title of Table
\label{table:1}      % is used to refer this table in the text
\centering                          % used for centering table
\begin{tabular}{c c c c}        % centered columns (4 columns)
\hline\hline                 % inserts double horizontal lines
Simulation  & Viscosity   & Disc & Planet-Star \\    
  name    & $\alpha$ & dim.       & initial distance (au) \\
\hline                        % inserts single horizontal line
  
   $M_{\rm 3D}$ & 0 & 3D & 5.2  \\ % inserting body of the table 
   $M_{{\rm 3D}\alpha 5}$ & $10^{-5}$ & 3D & 5.2 \\ 
   $M_{{\rm 3D}\alpha 4}$ & $10^{-4}$ & 3D & 5.2 \\ 
   $M_{{\rm 3D}\alpha 3}$ & $10^{-3}$ & 3D & 5.2 \\ 
   $M_{\rm 3D-13au}$ & 0 & 3D & 13  \\ 
   $M_{\rm 2D}$ & 0 & 2D  & 5.2 \\
   $M_{\rm 2D-13au}$ & 0 & 2D & 13 \\
   $M_{\rm 2D}$sg & 0  & 2D & 5.2 \\
   $M_{\rm 2D-13au}$sg & 0 & 2D & 13  \\
   
\hline                                   %inserts single line
\end{tabular}
\end{table}

\begin{figure}
\centering
   \includegraphics[width=\hsize]{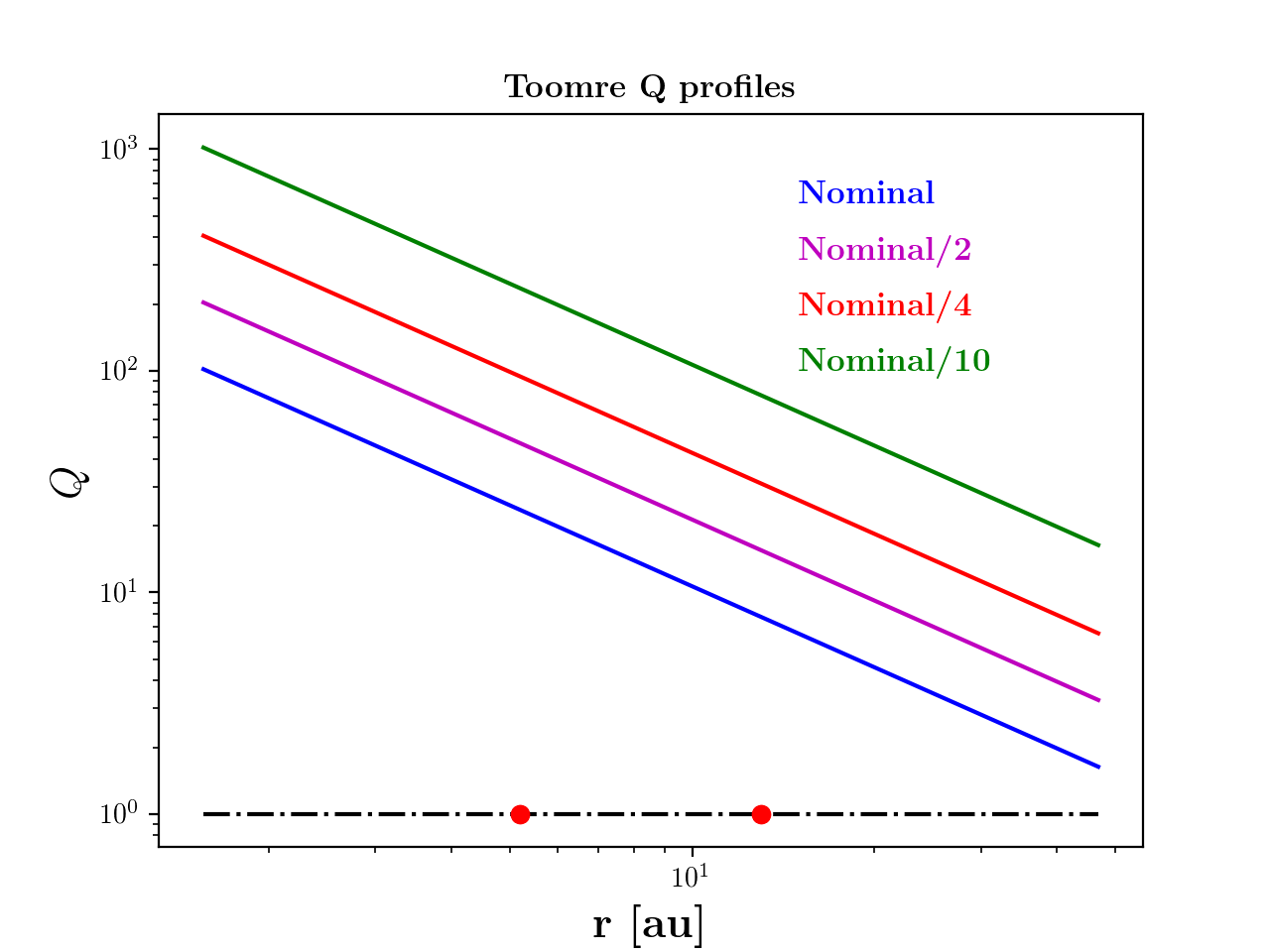}
      \caption{Toomre $Q$ values for 2D self-gravitating disc models. We focus on the nominal or $M$
      disc, but we also run control simulations with the mass of the disc divided by 2, 4 and 10. The $Q=1$ value is indicated by the dot-dashed line, and the locations of the planets at $r=5.2$ and $r=13$~au are indicated by the filled red circles.}
         \label{Fig:Qvalue} 
\end{figure}

\begin{figure*}
\centering
   \includegraphics[width=0.95\textwidth]{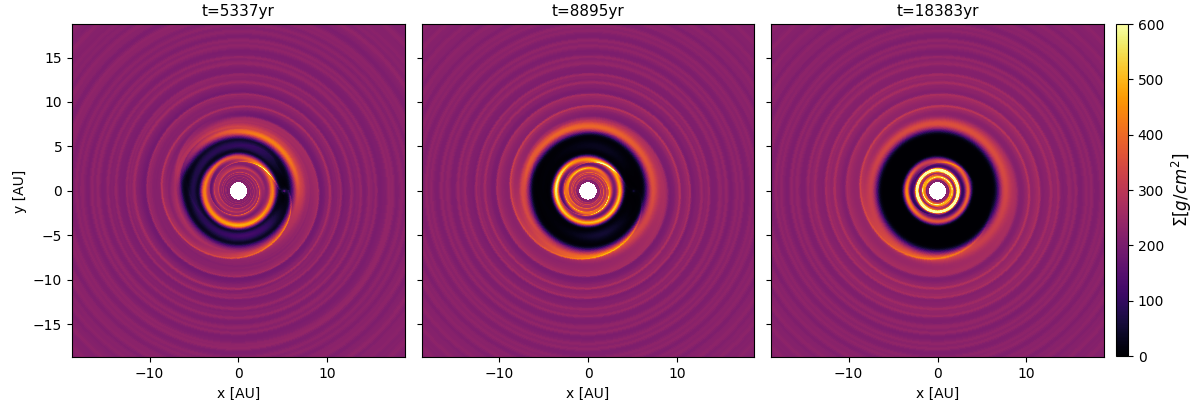}
   \includegraphics[width=0.95\textwidth]{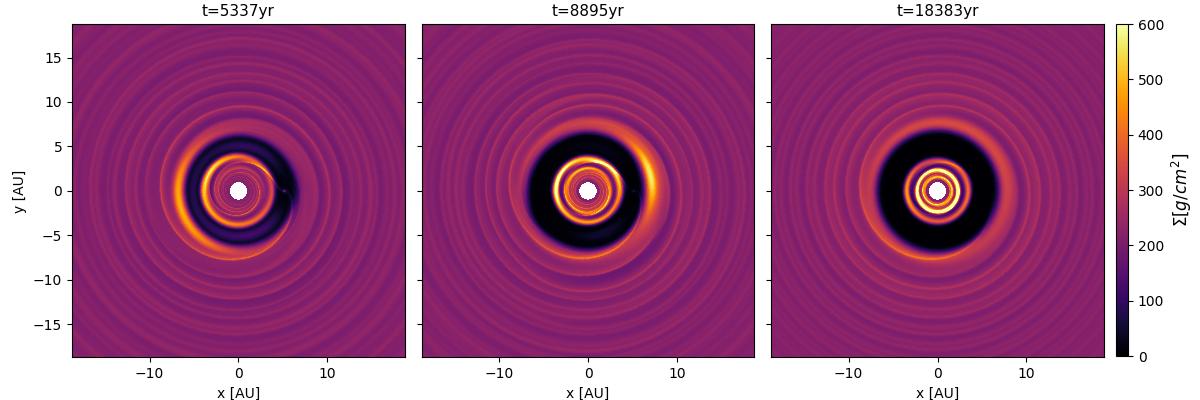}
   \includegraphics[width=0.95\textwidth]{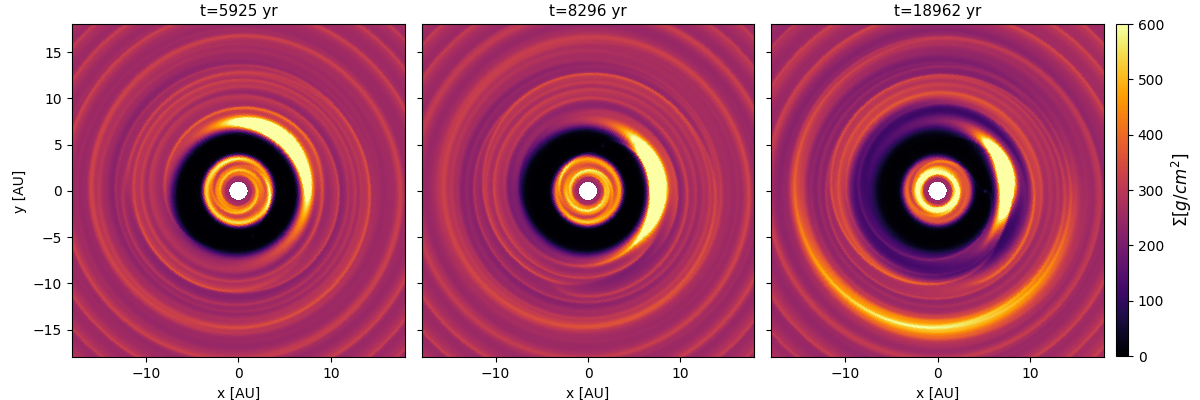}
   \caption{Top panels: Contours of surface density for the $M_{\rm 2D}$sg simulation of the nominal disc, for times shown (in years) at the top of each panel. Middle panels: Same as top panels but for the $M_{\rm 2D}$ simulation (i.e. without self-gravity).  Bottom panels: Same as top panels but for the $M_{\rm 3D}$ simulation. The planet is located at $r=5.2$~au.}
   \label{Fig:NominalSG-NOSG-r=1}
\end{figure*}

\subsection{Planet growth on fixed orbits}
In the 2D simulations the planet grows to its final mass over 800 orbits while being held on a fixed circular orbit. In the migration runs, the planet is released after the system has been evolved for a further 400 orbits. In the 3D runs the planet  grows to its final mass over about 200 orbits, and the system is evolved for a further 600 orbits before the planet is released in the migration runs. In this way we avoid the excitation of instabilities that would arise if the planet were initialised with its final mass (see also \citet{2020MNRAS.491.5759H}). 
%We  then run  the code for an additional 600 and PUT THE 2D VALUE orbits respectively in 3D and in 2D to let the disc relax to the presence of the planet before starting the migration phase. \par

\section{Planets on fixed orbits: comparing disc structures for simulations in 2D (with and without self-gravity) and in 3D}
\label{sec:2DSG-NOSG}
As discussed in Sect.~\ref{sec:Intro}, \citet{ZhuBaruteau16} have shown that the evolution of a protoplanetary disc with a vortex at a pressure bump depends on whether the indirect term and self-gravity are accounted for. Neglecting both of them is appropriate only for a disc with very low mass and results in a single vortex that remains at the location of the pressure bump. For more massive discs, the inclusion of the indirect term, without self-gravity, results in a large, radially-extended vortex that migrates inwards because of the excitation of strong spiral density waves. This leads to significant global restructuring of the disc. Including also self-gravity results in a more stable evolution, with one or more vortices forming at the pressure bump and remaining there for the duration of the simulation.

\begin{figure*}
    \centering
    \includegraphics[width=0.32\textwidth]{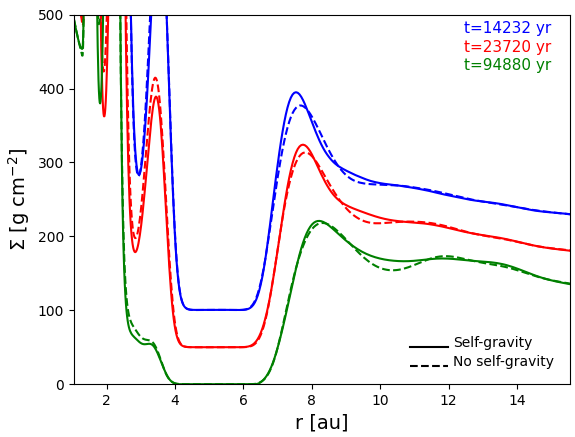}
    \includegraphics[width=0.32\textwidth]{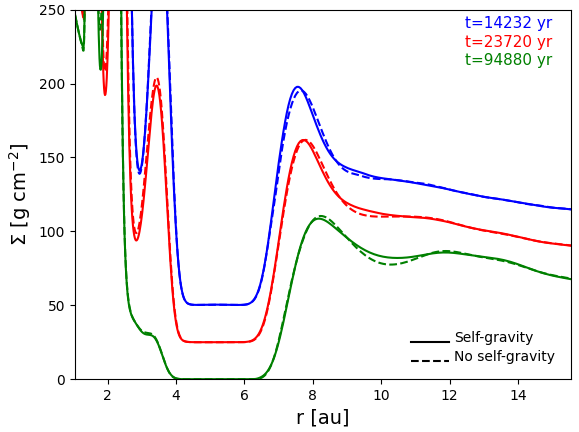}
    \includegraphics[width=0.32\textwidth]{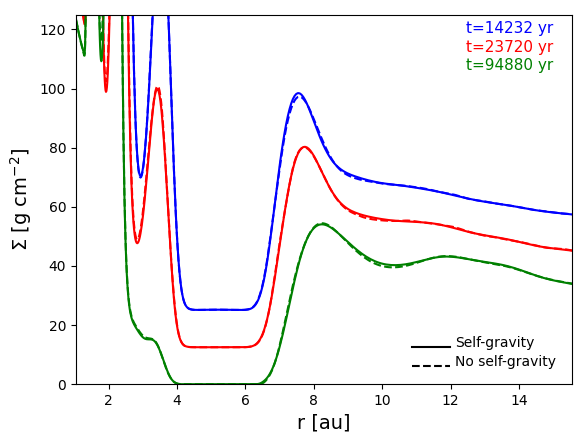}
    \caption{Azimuthally-averaged radial surface density profiles versus time for 2D runs with and without self-gravity, and a Jupiter-mass planet at 5.2~au. Moving from left to right: simulations $M$, $M/2$ and $M/4$ (i.e. decreasing disc masses). Note that the curves corresponding to different times have been vertically offset for clarity.}
    \label{fig:surfdenPlanet5.2AU}
\end{figure*}

The formation of a gap-forming planet in an inviscid protoplanetary disc inevitably leads to the formation of a vortex at the pressure bump located at the outer edge of the gap, and the subsequent evolution is expected to depend on the disc mass because of the influence of the indirect term and self-gravity. In this section, we first present results from a suite of 2D simulations that specifically address the question: \emph{under which conditions does self-gravity become unimportant?}. Then, for the conditions for which self-gravity is not important we present 3D simulations without self-gravity. We note that all simulations include the indirect term.

\citet{ZhuBaruteau16} suggested that a controlling factor is the value of the Toomre $Q$ parameter at the vortex location. The radial profile of $Q$ is shown in fig.~\ref{Fig:Qvalue} for different disc models: $M$, $M/2$, $M/4$ and $M/10$, where the disc mass is nominal or divided by 2, 4 and 10 respectively. Because $Q$ depends on radius, we consider models in which the giant planet is located at either $r_{\rm p}=5.2$ or $r_{\rm p}=13$~au.  We discuss in the following the case of a planet at $r_{\rm p}=5.2$ and we dedicate section \ref{sec:FarPlanets} to a discussion about planets forming farther out from the star ($r_{\rm p}=13$ au). Our main suite of 2D and 3D simulations adopt a local cooling time of $\tau_{\rm c}=1$~orbital period. We have found, however, that qualitative changes in the results (especially in 2D) arise when the cooling time changes, and hence we present a brief analysis of how the cooling time affects the results in Appendix~\ref{sec:Appendix1}.

\subsection{Planet at $r_{\rm p}=5.2$ au, 2D model}
\label{sec:r1}
We  consider the disc evolution with the planet orbiting at $r_{\rm p}=5.2$~au using the nominal parameters described in Sect.~\ref{sec:3.1}. We note the local Toomre parameter $Q(r=5.2~{\rm au})=23.5$, the aspect ratio $h(r=5.2~{\rm au})=0.05$ and the local cooling time $\tau_{\rm c}=1$~orbital period.

Our discussion will focus on the part of the disc that orbits exterior to the planet. During the early stages, the simulations show the formation of a gap and of 3 or 4 vortices that quickly merge into a single vortex at the outer edge of the gap. 
%*** I WOULD REMOVE THE FOLLOWING BECAUSE NOT REALLY TRUE IF THE PLANET IS AT 5 AU The evolution after this time depends strongly on the disc mass, whether or not self-gravity is included, on the location of the planet, on the equation of state and on the dimension of the problem.***

Results for the nominal $M_{\rm 2D}$sg and $M_{\rm 2D}$ simulations are shown in fig.~\ref{Fig:NominalSG-NOSG-r=1}. The self-gravitating simulation produces a vortex at the gap outer edge, which digs a shallow secondary gap at about 10~au by the emission of spiral waves, and weakens over time. Fig.~\ref{fig:surfdenPlanet5.2AU} shows that the size of the pressure bump slowly decreases over time. The non-self-gravitating simulation produces a stronger vortex and a deeper secondary gap than in $M_{\rm 2D}$sg, but the vortex is also observed to weaken over time and in the rightmost panels in Fig.~\ref{Fig:NominalSG-NOSG-r=1} we see that the final discs have very similar structures. Comparing self-gravitating and non-self-gravitating models, the radial profiles of the azimuthally averaged surface densities are seen to differ in detail in Fig.~\ref{fig:surfdenPlanet5.2AU}, mostly concerning the depth of the secondary gap at $\sim 10$~au, but are qualitatively similar. In both the self-gravitating and the non-self-gravitating simulations, the outer edge of the main gap is found to become Rayleigh unstable after approximately 800 planet orbits ($\sim 9,500$ yrs) and this coincides with the dissipation of the vortex, suggesting a  causal link.

We have checked (Fig.~\ref{fig:surfdenPlanet5.2AU}, middle and right panels) that, when we reduce the disc mass by factors of 2 and 4, the self-gravitating and non-self-gravitating models show rapid convergence in their behaviour even concerning small details, such as the depth of the secondary gap. We conclude that for the chosen parameters, these models all show good convergence in their behaviour regardless of whether self-gravity is included or not.

\subsection{Planet at $r_{\rm p}=5.2$ au, 3D model}
The 2D approximation is commonly used since protoplanetary discs are vertically thin and very often 2D results are shown to be consistent with results obtained with 3D models. If this is true in viscous discs, caution has to be taken for inviscid discs. Results for the $M_{3D}$ simulation are presented in Fig.\ref{Fig:NominalSG-NOSG-r=1}, bottom panels. We notice that a vortex forms at the outer edge of the gap also in 3D, and that it is radially larger and more massive than the vortex formed in the 2D models. Vortex exchange of angular momentum with the disc results in the evolution of the gap's outer edge. 
On the timescale of Fig.\ref{Fig:NominalSG-NOSG-r=1},  the vortex appears to be persistent over time, though slightly smoothing out.
The vortex exerts a pressure wave on the disc which superposes and interferes (sometimes non linearly) with the wake launched by the planet, which explains why the wave pattern observed in Fig.\ref{Fig:NominalSG-NOSG-r=1} (bottom panels) is more complex than that observed in 2D simulations. Through the wave the vortex exerts a torque on the disc which is effective enough to open a secondary gap   (Fig.\ref{Fig:NominalSG-NOSG-r=1}, bottom right panel) that is more pronounced than in the equivalent 2D simulation. 

\begin{figure*}
    \centering
    \includegraphics[width=0.45\textwidth]{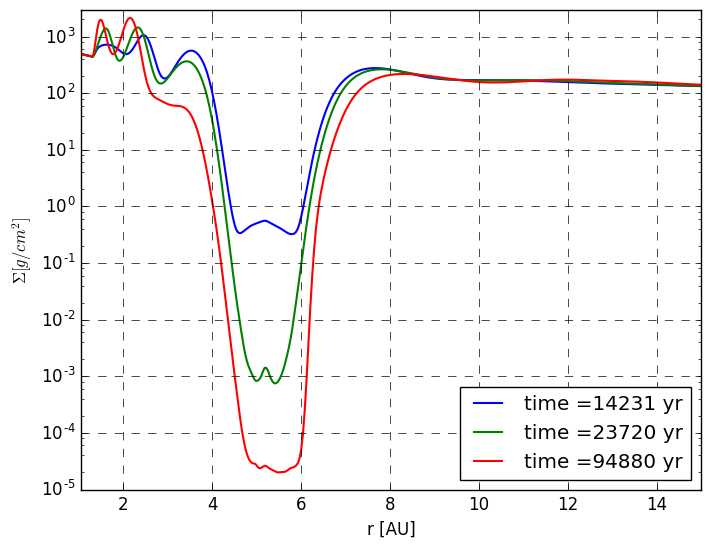}
    \includegraphics[width=0.45\textwidth]{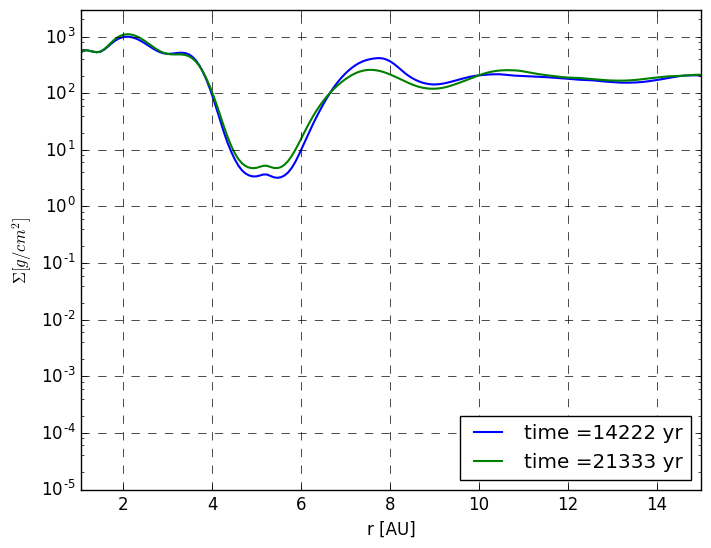}
      \caption{Azimuthally-averaged radial surface density profiles at different times for nominal disc mass ($M$) simulations with a Jupiter-mass planet at 5.2~au. Left panel:
      Same as the left panel of Fig. \ref{fig:surfdenPlanet5.2AU} for simulation $M_{\rm 2D}$, except that the scale on the $y$-axis is logarithmic, to appreciate the gap's depth and its evolution with time. Right panel: the same for the $M_{\rm 3D}$ simulation. The gaps in 2D and 3D inviscid discs (with adiabatic EoS and a cooling timescale equal to one planet's orbital period) appear very different; see text for discussion.}
    \label{fig:surfdenPlanet5.2AUComparison}
\end{figure*}

When looking at the radial surface density profiles in Fig.\ref{fig:surfdenPlanet5.2AUComparison}, we see that the planet-induced gaps appear very different. In 2D the gap becomes deeper and wider with time.  In 3D the gap looks almost stationary over time; the gap is even partially refilled and, at time $t=23720 \, \rm{yr}$, contains 3 orders of magnitude more gas than in the 2D simulations. 
This difference between 2D and 3D gaps in low viscosity discs was already shown in \citet{Morbidellietal14}. In 3D the  gap's depth is regulated by the gas' meridional circulation which continuously refills the gap, so that the gap's depth saturates at a much larger value than in 2D simulations. Consequently the disc does not satisfy Rayleigh's criterion for instability and the vortex is not smoothed out by the diffusion that this instability generates. In turn, the vortex exerts a negative torque on the outer edge of the gap, which contributes significantly to limiting the gap's depth. The difference between 2D and 3D simulations persists also for smaller disk masses. However, Appendix~\ref{sec:Appendix1} will show that the results of 2D simulations will approach those of 3D simulations when shorter cooling times are assumed, e.g. $\tau_{\rm c}=0.01$~orbit.

\begin{figure}
\centering
       \includegraphics[width=\hsize]{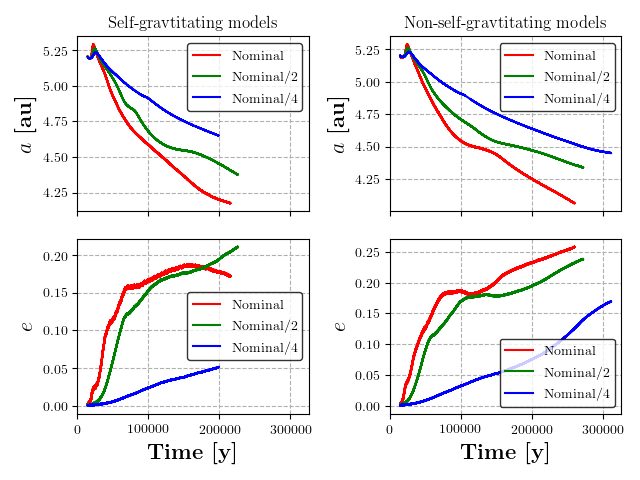}
          \caption{Left panels: Evolution of semi-major axes and eccentricities for planets in the simulations $M_{\rm 2D}$sg, $M_{\rm 2D}/2$sg and $M_{\rm 2D}/4$sg. Right panels: Same as left panels except for the corresponding non-self-gravitating disc models. }
         \label{Fig:aemigrM2DPl5.2au}
   \end{figure}
   
\section{Planet migration}
\label{sec:migrr1}
We investigate the migration of the planet initially at $r=5.2$~au, for which we have shown that self-gravity  has essentially no impact even for the nominal disc mass and $\tau_c=1$ orbit. We have shown that 2D and 3D discs behave very differently for these model parameters; it is therefore instructive to study planet migration in both cases in order to assess the range of behaviours that can occur in inviscid disc models even though, in principle, the 3D case should be more realistic than the 2D one. 

In order to better understand what follows, remember that a giant planet opening a gap in the disc can migrate only at the speed at which the disc can readjust itself in order for planet and gap to migrate together \citep{Ward1997}. In viscous discs, the gas can displace radially on the viscous timescale, which sets the planet migration speed to be proportional to $\nu/r$. In the inviscid case, the gap can move radially only in two ways. One possibility is that the planet displaces gas from the inner edge to the outer edge of the gap \citep{Duffelletal14, Durmann and Kley 2015}. \citet{Robertetal18} showed that this is not possible in the limit of low viscosity if the planet is on a circular orbit, because the gap is too wide with respect to the location of the separatrices of the planet's horseshoe-shaped coorbital region. But, if the planet acquires a sufficiently eccentric orbit, it can transfer gas from the inner to the outer disc. Then it can migrate inwards, irrespective of the viscous timescale. The other possibility is that the vortex forming at the outer edge of the gap pushes gas into the gap or migrates itself inwards due to the pressure waves that it generates in the disc \citep{Paardekooperetal10, ZhuBaruteau16}. In this case the outer disc can spread inwards on a timescale that is not related to viscous relaxation but is that of vortex-disc interactions (unrelated to viscosity). Thus, the planet can migrate on the same timescale, while pushing the inner disc inward. As we show below, 2D and 3D simulations display these two mechanisms.

\subsection{2D model}
\label{migr2D}
\begin{figure}
\centering
       \includegraphics[width=\hsize]{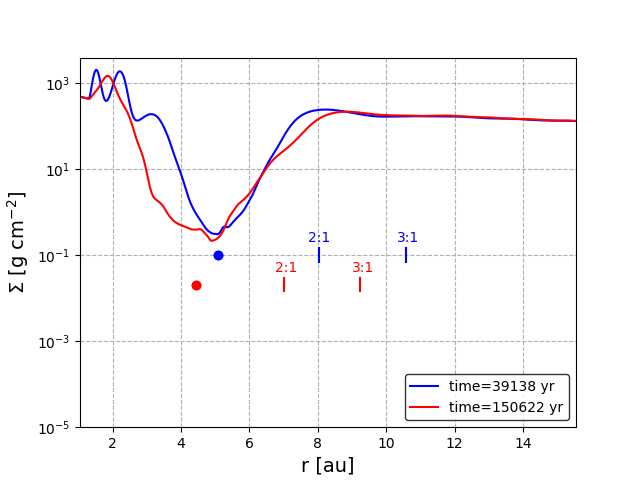}
          \caption{Surface density profile at the times shown in the legend for the run $M_{\rm 2D}$. Also shown are the semi-major axes of the planet (filled circles), and the locations of the 2:1 and 3:1 outer Lindblad resonances at the times indicated.}
         \label{Fig:surfacedensityM2DPl5.2au}
   \end{figure}
The evolution of the semi-major axes and the eccentricities are shown in Fig.~\ref{Fig:aemigrM2DPl5.2au} for the self-gravitating and non-self-gravitating $M$, $M/2$ and $M/4$ models. We begin by noting that the evolution of these quantities is similar in the self-gravitating and non-self-gravitating runs, which supports the conclusion of section~\ref{sec:r1} that self-gravity is unimportant in this disc model. From now on we just discuss in detail the results of the non-self-gravitating nominal model in this section. 

Fig.~\ref{Fig:aemigrM2DPl5.2au} shows that after an initial adjustment phase when the planet is released from its fixed circular orbit, the planet migrates inwards at an approximately steady rate of $\sim 5.55$~au/Myr. During this time the eccentricity rises steadily up to $e_{\rm p}\sim 0.25$. As discussed in section~\ref{sec:r1}, a vortex forms at the outer gap edge as the planet grows and opens a gap, but the vortex is short lived and dissipates shortly before the planet is released. Hence, the planet is embedded in a disc in which there is essentially no internal angular momentum transport, except due to the density waves excited by the planet itself.

A gap-forming planet in a \emph{viscous} disc can experience eccentricity growth when the gap becomes deep enough that eccentricity driving by external Lindblad resonances (ELRs) dominates the combined eccentricity damping due to coorbital Lindblad resonances (CLRs, which sit at the location of the planet's semi-major axis, $a_p$) and corotation resonances (CORs) \citep{GoldTre1980}. Diagrams showing the locations of the various competing resonances are shown in \citet{GoldSari03}. To first order in the planet eccentricity, $e_p$, and considering the outer disc only, we note that the 1:3 ELR (located at $r=2.08 a_p$) is the outermost eccentricity driving resonance, and the 1:2 COR (located at $1.58 a_p$) is the outermost damping resonance. Growth from an initially very small (formally infinitesimal) eccentricity can occur when the gap is wide enough that it extends beyond the 1:2 COR, such that the 1:3 ELR can drive up the eccentricity without competition from CORs \citep{PapaloizouNelsonMasset2001}. Alternatively, eccentricity growth can occur when the initial eccentricity has a significant finite amplitude, because the CORs can then become partially saturated, and hence unable to dominate the ELRs \citep{GoldSari03, DuffelChiang15}. 

In an \emph{inviscid} disc, with no internal transport of angular momentum, CORs  saturate completely and the formation of a deep gap  renders the CLRs ineffective. Thus eccentricity growth should be expected, as shown in the simulations. Fig.~\ref{Fig:surfacedensityM2DPl5.2au} depicts the surface density profile in the disc at two different times; also indicated are the planet semi-major axes and the locations of the 1:2 COR and the 1:3 ELR.  At $t=39138$~yr, when the eccentricity has already undergone significant growth, the gap is insufficiently wide to deplete the gas at the 1:2 resonance, suggesting this resonance is saturated due to the absence of viscosity or any other source of angular momentum transport such as Reynolds stress caused by a vortex.

The growth of the eccentricity gives rise to the high-eccentricity mode of migration discussed above. It is clear from comparing Figs.~\ref{Fig:surfacedensityM2DPl5.2au} and \ref{fig:surfdenPlanet5.2AUComparison} that the gap structure evolves very differently when the planet's orbit is free to change versus when it is kept fixed and circular. In particular, the gap becomes less deep but wider as the planet moves away from the original outer edge of the gap while at the same time becoming eccentric. The eccentric orbit induces a flow of gas from the inner to the outer disc, and this appears to be sufficient to allow the planet to migrate inwards at the rate indicated by Fig.\ref{Fig:aemigrM2DPl5.2au} 
(i.e. $\sim 6.7$~au per Myr) over large radial scales and long time scales. It is noteworthy that the maximum eccentricity of
$e \sim 0.25$ obtained by the planet is much larger than that reported by \citet{DuffelChiang15}, who demonstrated that a Jovian-mass planet in a viscous disc with initial eccentricity $e=0.04$ could experience growth only up $e=0.07$ before interaction with the gap edges stalls the eccentricity growth. Here, it appears that a Jovian mass planet in an inviscid disc can open a wider gap because torques exerted on the disc at ELRs as the eccentricity grows operate unopposed by viscosity. Hence, a larger eccentricity can develop.

\subsection{3D model}
\label{sec:3DNominal}
\begin{figure}
\centering
       \includegraphics[width=\hsize]{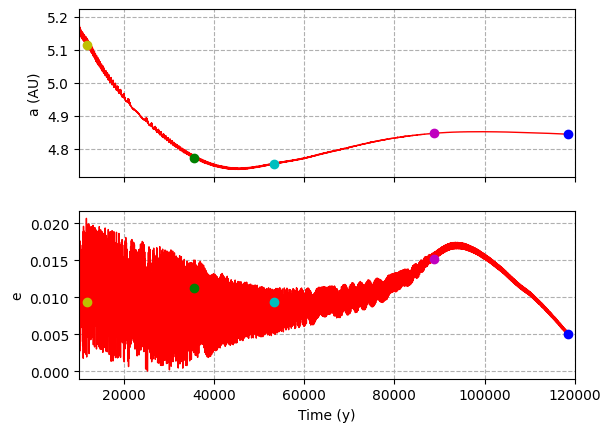}
          \caption{Evolution  of semi-major axis and eccentricity for a migrating planet in   simulation M$_{\rm 3D}$. Notice the outward migration after 45,000y. The colored dots give the values of $(a,e)$ at times corresponding to the panels of Fig.\ref{Fig:Nominal3DMigrPl5.2au}. }
         \label{Fig:aemigrM3DPl5.2au}
   \end{figure}

\begin{figure*}
\centering
    \includegraphics[width=0.95\textwidth]{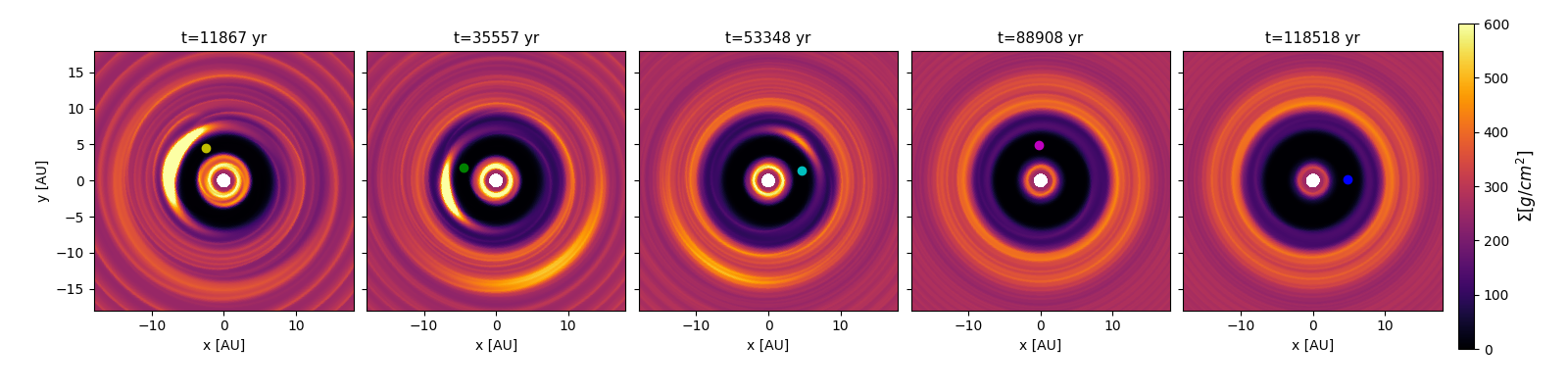}
   \caption{Contours of surface density for the $M_{\rm 3D}$ simulation and a migrating Jupiter-mass planet. The panels correspond to different times, reported on top of each panel, corresponding to the dots in Fig.~\ref{Fig:aemigrM3DPl5.2au}. It appears clearly that a secondary gap is carved by vortex-disc interaction (second and third panels) and that the vortex weakens by spreading radially, leaving the outer disc partially depleted (fourth and fifth panels).}
   \label{Fig:Nominal3DMigrPl5.2au}
\end{figure*}

The planet is released after 800 orbits at $r=5.2$~au, i.e. at the time corresponding approximately to the middle bottom panel in fig~\ref{Fig:NominalSG-NOSG-r=1}. The planet remains on an orbit with small eccentricity ($e\sim 0.01$) because (i) the gap is not very deep and (ii) the CORs remain unsaturated because of the meridional circulation and the presence of the vortex, which generates an effective viscosity by exciting Reynolds stresses. Because of the small planet eccentricity, the passage of gas through the gap is virtually null. Thus, the migration mode observed in the previous 2D simulations is not observed. Nevertheless, the planet migrates inwards during the first $\sim 40,000$~yr (Fig.~\ref{Fig:aemigrM3DPl5.2au}). Migration is possible because the vortex itself tends to migrate inwards, thus refilling the gap as the planet moves away. Vortex migration is due to its ability to exchange angular momentum with the disc, via the generation of spiral density waves \citep{Paardekooperetal10}. Vortex migration is inwards because the vortex is located at the outer edge of the gap, so that the disc inside the vortex's orbit is strongly depleted and the interaction with the outer disc dominates. By the action-reaction principle, the vortex transfers angular momentum to the disc beyond its orbit, carving a
secondary gap (Fig.\ref{Fig:Nominal3DMigrPl5.2au}). Once this gap is formed, the vortex slows down, spreads radially and eventually dissipates completely, leaving the outer disc partially depleted near the planet's orbit. 

\begin{figure}
\centering
       \includegraphics[width=\hsize,height=10truecm]{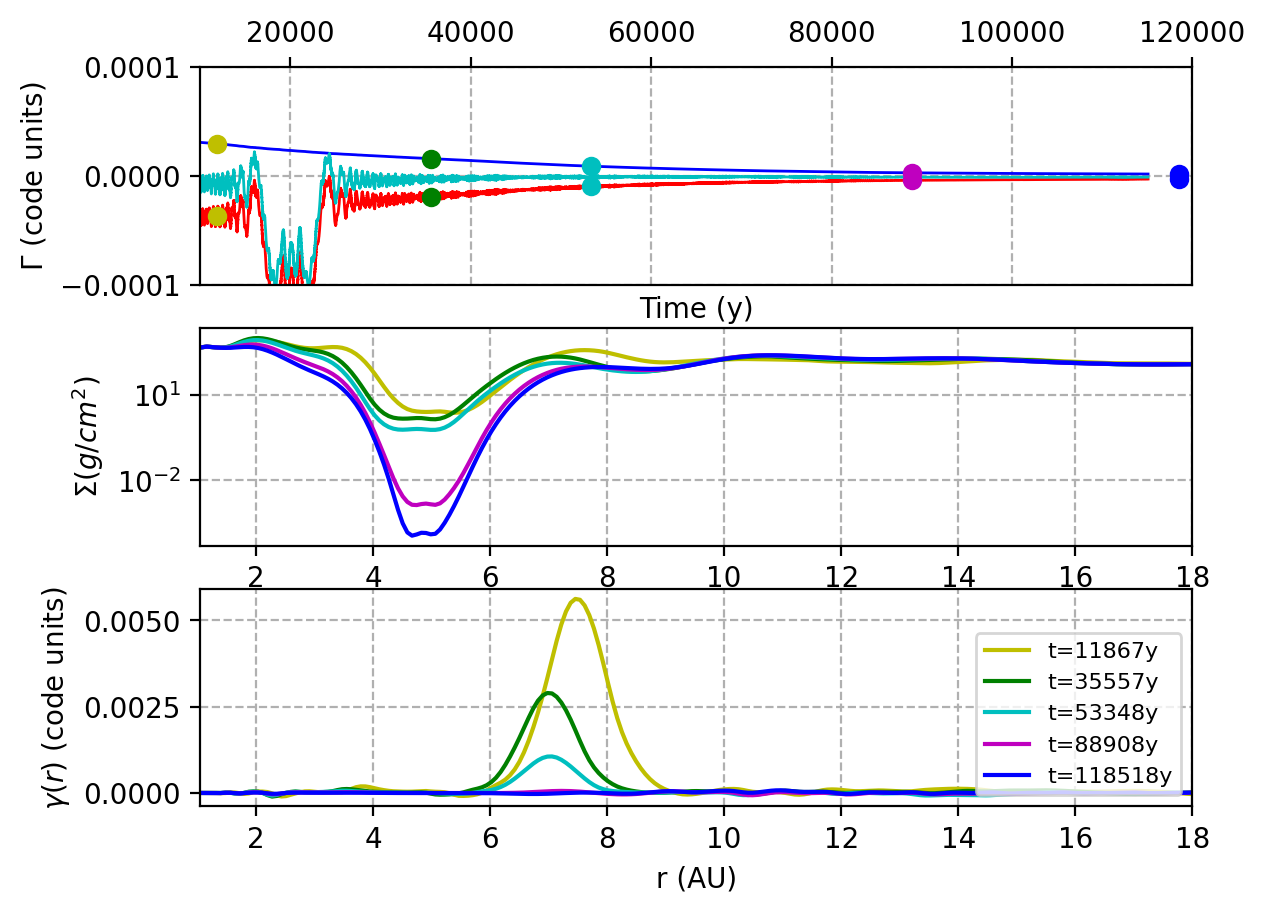}
          \caption{Top panel: evolution  of specific torques  from the inner (blue curve) and outer disc (red curve) and of the total specific torque (cyan) in simulation $M_{\rm 3D}$. A running average has been applied over 500 orbits to smooth out the oscillations of the torque exerted by the outer disc. Such oscillations are due the synodic period of the vortex with respect to the planet. At times corresponding to the  dots in the top panel (the same times selected in Fig.~\ref{Fig:aemigrM3DPl5.2au} and Fig.\ref{Fig:Nominal3DMigrPl5.2au})
          we plot: (middle panel) surface density profiles;
          (bottom panel) { specific} radial torque  density $\gamma (r)$ { (see Eq.\ref{radialtorquedens} in the text)}.}
         \label{Fig:tqandsurfdensPl5.2au}
   \end{figure}

The depletion of the outer disc relative to the inner disc creates an imbalance in the torques exerted on the planet, with the positive torque exerted by the inner disc becoming the (slightly) dominant one (Fig.~\ref{Fig:tqandsurfdensPl5.2au}, top panel). As a consequence, the direction of migration is reversed and the planet slowly migrates outwards (Fig.\ref{Fig:aemigrM3DPl5.2au}, top).  However, the process of outward migration cannot continue indefinitely. The torque contribution from the inner disc decreases when the planet moves outwards (because its distance from the inner disc increases), eventually balancing out with the negative torque exerted by the outer disc. Consequently, migration slowly stops and the planet remains on a quasi-circular orbit embedded in a gap that is much wider than the initial one, resulting from the merging of the planet's main gap with the secondary gap carved by the now-disappeared vortex (Fig.\ref{Fig:Nominal3DMigrPl5.2au}, rightmost panel). 
Incidentally, this breaks the relationship  usually used to deduce the planet's mass from the gap's width  in observed disks.
 Fig.\ref{Fig:tqandsurfdensPl5.2au}, central panel shows that when the planet is at rest the gap becomes much deeper than the one during the migration phase. Nevertheless, enough gas remains in the gap and the CORs are sufficiently desaturated by the meridional circulation to damp the eccentricity of the planet once it stops migrating (Fig.\ref{Fig:aemigrM3DPl5.2au}, bottom).

In order to study more precisely the contributions of different regions of the disc - and of the vortex in particular - to planet migration, we show in Fig.\ref{Fig:tqandsurfdensPl5.2au} (bottom panel) the { specific} torque $\gamma(r)$   exerted on the planet by a ring of the disc placed at $r$ with width $dr$ which is:
\begin{equation}
    \gamma(r){\rm d}r = \int _{-\pi/2}^{\pi/2} \int_{0}^{2\pi} \left(\frac {G\rho(r,\theta,\phi)r^2sin(\theta)(\vec r -\vec r_p)} {(r-r_p)^3} \wedge \vec r {\rm d} r\right)_z{\rm d}\theta {\rm d}\phi \ .
    \label{radialtorquedens}
\end{equation}
{ where the subscript z indicates the vertical component of the vector enclosed in the brackets.}
The integral of $\gamma(r)$ over the radial coordinate provides the total {specific}
torque $\Gamma$, plotted in the top panel of Fig.\ref{Fig:tqandsurfdensPl5.2au}:
\begin{equation}
\Gamma =\int _{r_{min}}^{r_{max}} \gamma(r){\rm d}r\ .
\label{totaltorque}
\end{equation}
  The vortex provides periodically positive to negative contributions to $\gamma$ depending on the planet-vortex azimuthal distance that average to zero over a planet-vortex synodic period.
To highlight the vortex location, its radial migration and its fading over time, it is
convenient to select snapshots at which  the planet and the vortex have the same relative shift in azimuth.  We did this by sampling the hydrodynamical quantities every 1/10 of orbit and choosing snapshots with the vortex always ahead of the planet's location as shown in Fig.\ref{Fig:Nominal3DMigrPl5.2au}, corresponding to the maximal  value of $\gamma$ at the vortex's radial location during the vortex's synodic period.
 The fact that $\gamma$ is positive in these snapshots  does not imply that  the vortex exerts a net positive torque on the planet (remember that  averaged over a synodic period the torque is zero).
With this procedure, the bottom panel of Fig.~\ref{Fig:tqandsurfdensPl5.2au} reveals that the position of the maximum in $\gamma$, corresponding to the vortex's radial position,
   is located  at the gap's outer edge (middle panel).
   Both vortex and gap's edge
   shift inwards with time, until t=35,557~yr. Moreover the maximum value of $\gamma$ decreases with time, revealing the weakening of the vortex. For times larger than t=88,000~yr  the vortex has been completely dissipated (2 rightmost panels of Fig.\ref{Fig:Nominal3DMigrPl5.2au}) and accordingly there  is no positive contribution to $\gamma$.

\par

\begin{figure}
\centering
    \includegraphics[width=\hsize]{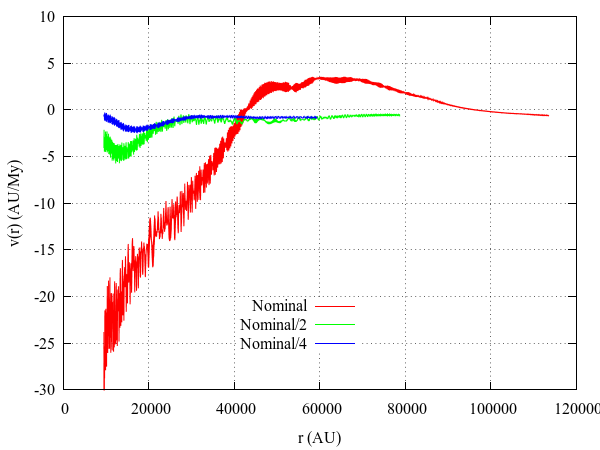}
   \caption{Migration speed as a function of time, for discs with different masses in 3D simulations. The planet always starts at 5.2~au. The migration speed decreases, as expected, with the disc's mass.  The direction of migration is reversed at $t \simeq 45,000 \rm{yr}$ in the nominal mass disc case (simulation $M_{\rm 3D}$).}
   \label{Fig:speed3D5.2au}
\end{figure}

In summary, after an initial short ($\sim$40,000~yr) phase of relatively fast inward migration, during which the planet migrates less than 0.5~au (Fig.~\ref{Fig:speed3D5.2au}), the migration of the planet slows down, then reverses direction, leading to an outward migration that naturally damps out at 4.85~au.   A test of convergence with respect to the resolution of the simulation is provided in the appendix.

When we decrease the mass of the disc, the vortex is less massive, roughly proportionally to the disc's surface density. However, the indirect term also scales with the disc mass and therefore we cannot expect a simple linear decrease of the migration speed with the disc's mass.  In the $M_{\rm 3D}/2$ and $M_{\rm 3D}/4$ simulations the planet migrates slowly inward at a speed of about 1 au/Myr (slightly faster for $M_{\rm 3D}/2$ than for $M_{\rm 3D}/4$). In both cases we observe the vortex spreading radially and dissipating with time while opening  a small dip beyond the vortex's orbit. The torque imbalance between inner and outer disc appears to be not enough for the reversal of the migration direction. 

\subsection{Definition of vortex-driven migration}
\label{Def}
The result illustrated above shows that, in inviscid disks, migration of giant planets on quasi-circular orbit is sustained because the migration of the vortex allows to refill the gap left-behind by the migrating planet. We call this process ``vortex-driven'' migration, to distinguish it from the ``vortex-induced'' migration of  \citet{LinPapaloizou10,LinPapaloizou11} which was due to the passage of the vortex through the planet's coorbital region. As we mentioned in the introduction, vortex migration is much less prominent if self-gravity is taken into account \citep{ZhuBaruteau16}. Indeed, in Appendix~\ref{sec:Appendix2D} we will show a 2D simulation with self-gravity and fast disc cooling that will behave similarly to our $M_{\rm 3D}$ simulation, i.e. with a pronounced vortex and a slowly migrating planet on a quasi-circular orbit, but showing no vortex migration. In these cases, planet migration is possible because the vortex, although not moving, is able to push gas inwards, refilling the gap left-behind by the planet. We include also this case in the ``vortex-driven'' migration category. In both cases, the migration of the planet occurs on the timescale characterising the vortex-disc interaction (either manifested by vortex migration or vortex-driven disc spreading), which is unrelated to the disc's viscosity. Thus vortex-driven migration is conceptually different from Type-II migration. Vortex-driven migration is dominated by Type-II migration if the disc's viscosity is large enough (as quantified in the next section) or superseded by the eccentric migration mode (as seen in Sect.~\ref{migr2D}) if the planet's eccentricity becomes large enough to allow the passage of gas across the gap, unlocking the planet from the disc's evolution. 

In summary, vortex-driven migration is a slow, inward and smooth migration due to the presence of a vortex beyond the orbit of the planet. It is very different from vortex-induced migration, which is a violent process inducing a phase of fast migration. Importantly, vortex-driven migration does not seem to be indefinite. By modifying the disk around it (i.e. spreading the disk inwards, opening a secondary gap) the vortex eventually fades away and its effect stops. When this happens, a transient phase of outward migration is even possible if the disappearance of the vortex leaves a gap that is asymmetric relative to the position of the planet. 

Vortex-driven migration, being slow and transient, offers a new and interesting explanation of the existence of the cold-Jupiters, without requiring (as in \citet{ColemanNelson2014, Bitschetal2015}) that these planets formed much beyond the snowline.

\begin{figure}
\centering
    \includegraphics[width=\hsize]{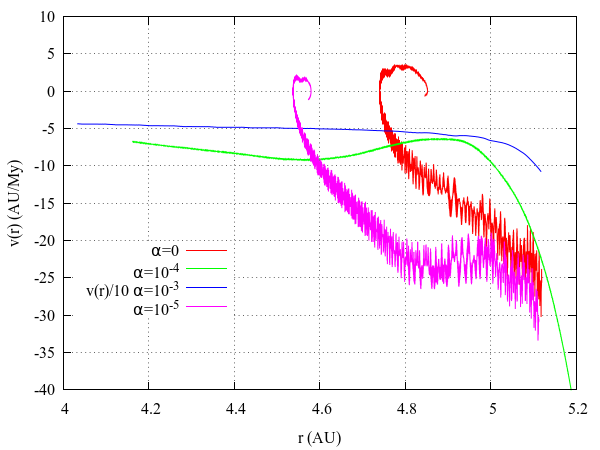}
   \caption{Migration speed as a function of planet's distance from the star in 3D simulations of discs with different viscosities (i.e. simulations $M_{\rm 3D}$, $M_{{\rm 3D}\alpha 5}$, $M_{{\rm 3D}\alpha 4}$ and $M_{{\rm 3D}\alpha 3}$; see Table 1).  A running average procedure has been used to reduce fluctuations, with the consequence  that all but the green curve start at about $5.1$ au instead of the released distance of $5.2$ au. In the $\alpha=10^{-3}$ and $10^{-4}$ cases migration is always inwards (negative radial speed).  Notice that we have divided by 10 the migration speed of $M_{{\rm 3D}\alpha 3}$ in order to scale it proportionally to viscosity relatively to the $\alpha=10^{-4}$ case. The migration pattern in the inviscid case does not respect this scaling and turns positive after a short phase of inward migration. The same happens in the $M_{{\rm 3D}\alpha 5}$ simulation. }
   \label{Fig:speed3D5.2auAlpha}
\end{figure}

\begin{figure*}
\centering
   \includegraphics[width=0.95\textwidth]{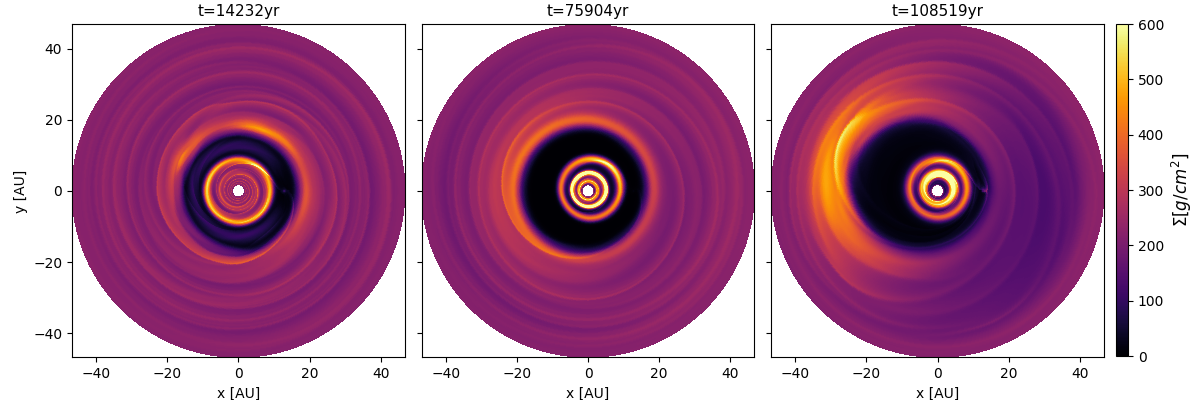}
   \includegraphics[width=0.95\textwidth]{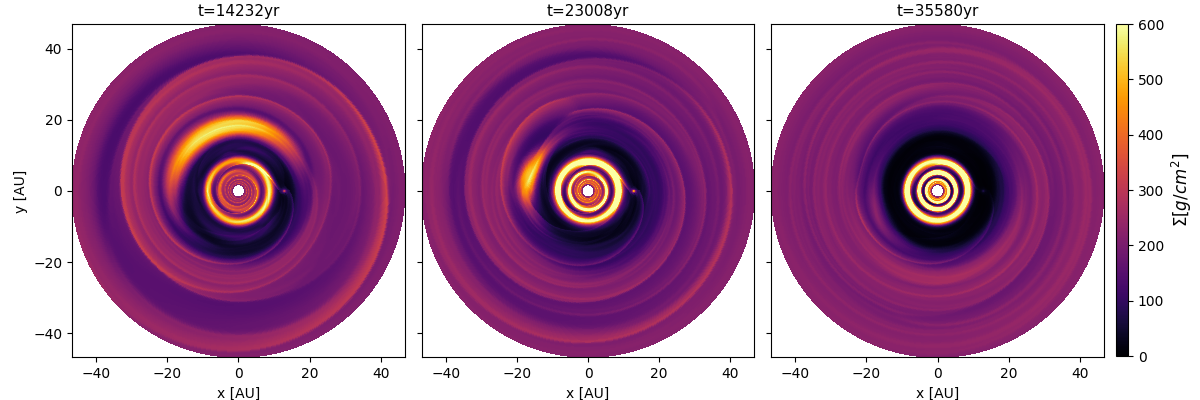}
    \includegraphics[width=0.95\textwidth]{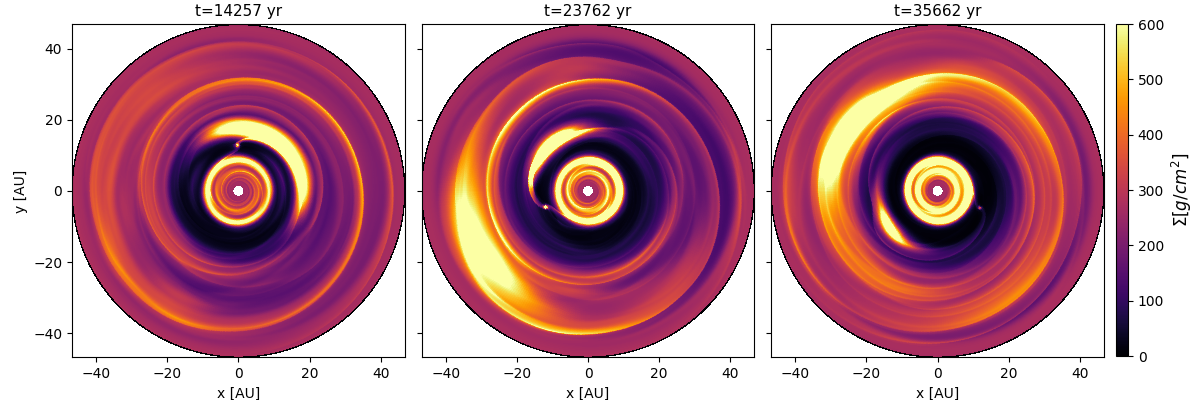}
   \caption{The same as Fig.~\ref{Fig:NominalSG-NOSG-r=1}, but for a planet located at $r_{\rm p}=13$~au.}
   \label{Fig:NominalSG-NOSG}
\end{figure*}

\subsection{Migration in viscous discs}
\label{sec:viscosity}
This section addresses the question: \emph{how low does the viscosity need to be for the vortex-driven migration to be dominant?} 
For this purpose we ran 3D simulations with values of the $\alpha$ viscosity parameter in the range $[10^{-5},10^{-3}]$.
In viscous discs the equation of state should consistently include a term corresponding to viscous heating. However, the viscous heating would also change the vertical disc structure
\citep{KBK09} making the comparison less clear. In order to compare equivalent discs we null the viscous heating contribution.

Fig.\ref{Fig:speed3D5.2auAlpha}
shows the migration speed as a function of the planet-star distance. Planets monotonically migrate towards the star for $\alpha$ down to $10^{-4}$, with a speed which scales with the viscosity, as expected in the Type-II migration regime. 
For $\alpha = 10^{-5}$ a vortex forms very similarly to the non-viscous case and vortex-driven migration is observed with speed slightly larger than in the non-viscous case and persisting for a longer time, thus driving the planet closer to the star (to $\sim 4.55$~au instead of $\sim 4.75$~au at the moment of migration reversal).
Therefore, we conclude that the transition between Type-II migration and vortex-driven migration occurs for values of  $\alpha$ in the interval $[10^{-5},10^{-4}]$. This makes vortex-driven migration an interesting process acting at viscosities that are not exceedingly small, but could be typical of realistic discs \citep{Turner2014}.

When considering non-viscous discs in numerical simulations there is always a non-zero numerical viscosity.  The fact that we see a different evolution for
decreasing values of $\alpha$ down to $10^{-5}$ (Fig.\ref{Fig:speed3D5.2auAlpha})
and that the simulation with $\alpha=10^{-5}$ is different from the inviscid simulation suggests that the numerical viscosity for our inviscid 3D simulation $M_{\rm 3D}$ is not larger than the one modeled assuming $\alpha=10^{-5}$. 

\section{Distant planets}
\label{sec:FarPlanets}

In some planet-evolution models, giant planets are postulated to start forming at large distances from the parent star. From the results of this paper, one could be tempted to conclude that in low-viscosity discs these planets remain at large distances throughout the disc's lifetime and may correspond to the giant planets observed by direct imaging surveys. Here, we warn that the results are not invariant with the distance of the planet from the central star. This is because the role of self-gravity becomes more prominent farther out.

\subsection{Fixed planet at $r_{\rm p}=13$ au, 2D model }
\label{sec:r2.5}
We start this discussion by considering 
 the  $M_{\rm 2D-13au}$ and $M_{\rm 2D-13au}$sg simulations.  The local Toomre parameter $Q(r=13~{\rm au})=7.74$ and the aspect ratio $h(r=13~{\rm au})=0.065$ for this model (see Fig.~\ref{Fig:Qvalue}). We plot in Fig.~\ref{Fig:NominalSG-NOSG}  contours of the surface density.  
\begin{figure*}
    \centering
    \includegraphics[width=0.32\textwidth]{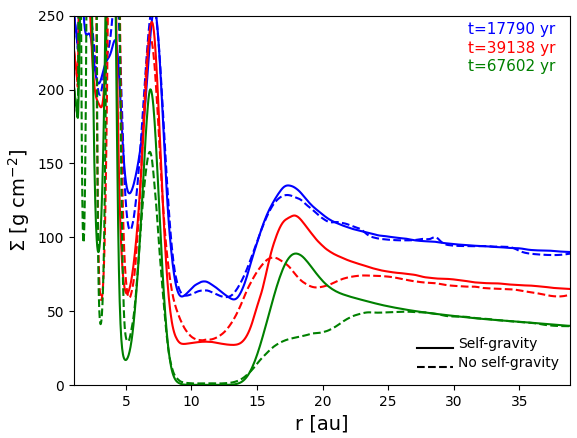}
    \includegraphics[width=0.32\textwidth]{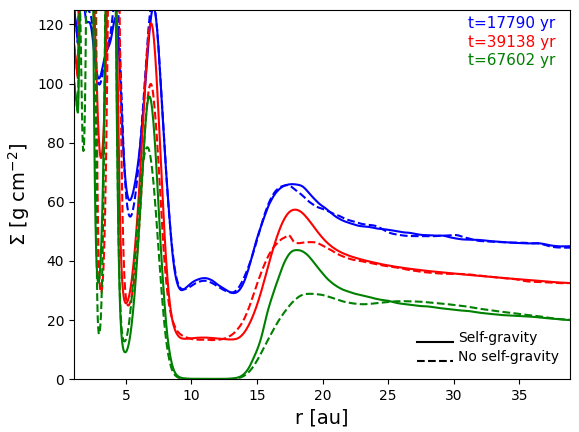}
    \includegraphics[width=0.32\textwidth]{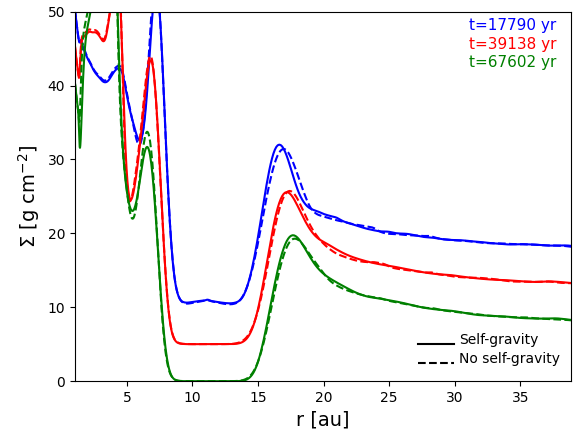}
    \caption{Azimuthally-averaged radial surface density profiles versus time for 2D runs with and without self-gravity, and a Jupiter-mass planet at 13 au. Moving from left to right: simulations $M/2$, $M/4$ and $M/10$.  Note that the curves corresponding to different times have been vertically offset for clarity.}
    \label{fig:surfdenPlanet13AU}
\end{figure*}

As a consequence of the relatively low $Q$ value at the planet's location, a behaviour similar to that described in \citet{ZhuBaruteau16} is replicated in our simulations. The self-gravitating disc ($M_{\rm 2D-13au}$sg) forms a single vortex that remains stably orbiting at the planet-induced pressure bump throughout the simulation and, although a detailed inspection of the surface density evolution shows some intermittency in the structure of the vortex and the pressure bump, the evolution is in general smooth and involves the vortex weakening with time. On longer time scales, however, the disc develops a strong $m=1$ eccentric mode, that eventually becomes so strong that it destroys the disc (remember that the planet is kept on a circular orbit, so that the disc's eccentricity cannot be limited by transferring the disc's angular momentum deficit to the planet). The non-self-gravitating run $M_{\rm 2D-13au}$ displays very different behaviour. A large vortex forms and migrates into the gap region and disperses, smoothing the profile of the gap outer edge. Over long run times (i.e. $>200,000$~yr, not shown in Fig.~\ref{Fig:NominalSG-NOSG}), we observe that non-axisymmetric structures form again in the disc. 

To reach genuine convergence between self-gravitating and non-self-gravitating disc in the case of distant planets we need to reduce the disc mass by a factor 10 (see Fig.~\ref{fig:surfdenPlanet13AU}), which brings $Q$ to $\sim 70$, comparable to the value in the $M/2$ simulations for $r_p=5.2$~au.  In this case the disc forms a single vortex at the outer edge of the gap that remains there over the simulation run time. The vortex weakens over time, and the only significant difference between the self-gravitating and the non-self-gravitating runs is the length of time required for vortex weakening to occur. 
Because of the lack of convergence in the disc behaviour for self-gravity and no-self-gravity runs we do not report a detailed analysis of migration, but it is worth mentioning that in the $M_{\rm 2D-13au}$sg simulation the planet becomes very eccentric (unsurprisingly, given the eccentricity acquired by the disc), which makes the planet go out of the gap at perihelion and aphelion, triggering very fast inward migration.

\subsection{Fixed Planet at $r_{\rm p}=13$ au, 3D model }
We remind the reader that our 3D code does not feature self-gravity. Thus, it is not a surprise that the disc's surface density (fig.~\ref{Fig:NominalSG-NOSG}, bottom panels) appears more similar to the 2D non-self-gravitating model (middle panels) than to the self-gravitating case (top panels). The vortex interacts with the disc and weakens while forming a secondary gap. At the outer edge of the secondary gap a second vortex forms while the inner vortex gets weaker. Reducing the disc mass by a factor 10, we have obtained results quite similar to the 2D models, with a weak vortex that appears to evolve very slowly over time. 

It is then clear that in order to study giant planet migration realistically at large distances in nominal discs one needs to use a 3D code with self-gravity, which to our knowledge has never be done before. Without self-gravity, only migration in low-density discs can be studied and we find a similar behavior as that discussed in section 5.2. 

\section{Discussion and conclusions}
In this paper we have examined giant planet migration in low viscosity discs with 2D and 3D numerical simulations. First, we have considered a Jupiter-mass planet at 5.2~au in a nominal disc { with  density at 5.2~au of 220 ${\rm g/cm^2}$ or smaller}. The presence of a giant planet in a low viscosity disc opens a gap and triggers vortex formation at the outer edge of the gap. The presence of the vortex raises the question of the importance of the self-gravity in the disc \citep{ZhuBaruteau16}. Thus, we compared 2D simulations with and without disc self-gravity to conclude that, apart for minor and temporary differences, the evolution observed  with a disc's cooling time $\tau_c=1$~orbit are equivalent to each other. Then we moved to 3D simulations of the planet-disc interaction. 3D simulations are in principle more realistic than 2D simulations because they can capture genuine 3D effects such as the meridional circulation of gas in the vicinity of gaps \citep{Morbidellietal14}, but our code FARGOCA does not allow to take self-gravity into account, so that its use is valid only where self-gravity is not relevant, as shown by the aforementioned 2D simulations. 

For the nominal model with $\tau_c=1$ orbit, the results of 3D simulations are quite different from those of 2D simulations: the gap is shallower and the vortex at its outer edge is much larger and persistent. The two aspects are linked in a positive feedback. Because the gap is shallower, the disc never becomes Rayleigh unstable, unlike in 2D simulations. Thus the vortex is not eroded by the diffusivity generated by this instability; in turn the presence of the vortex exerts a torque on the disc which helps to push gas into the gap. This effect, together with the meridional circulation of gas that exists only in 3D, makes the 3D gap shallower. 

As a consequence of the sharp differences in the gap's depth and vortex properties, we observe a big difference in the orbital evolution of the planet when the latter is free to migrate. In 2D simulations the planet rapidly acquires a significantly eccentric orbit with $e \sim 0.25$, apparently because the depth of the gap and the lack of effective viscosity operating in the disc allows coorbital and corotation resonances, that usually damp the eccentricity, to saturate \citep{GoldTre1980,GoldSari03}. 
In their paper, Goldreich and Sari conjectured that this mechanism can explain the large orbital eccentricities observed for extrasolar cold-Jupiters.  We find that this is unlikely. In fact, as soon as the planet's orbit becomes eccentric, the planet undergoes a relatively fast migration ($\sim 6.7$~au/Myr which, although slower than Type-II migration in an $\alpha=10^{-3}$ disk, implies full orbital decay in less than a Myr relative to the planet's initial position at 5.2~au).  Thus, we suggest it is unlikely that an eccentric giant planet could remain in the cold-Jupiter region at the end of the disc's lifetime, given the speed of migration. Moreover, we do not see such an eccentricity excitation in the 3D simulations, as discussed below. Thus, dynamical instabilities after the removal of the protoplanetary disc \citep{FordRasio2008,2008ApJ...686..580C,2008ApJ...686..603J} seem to provide a much more robust mechanism for the eccentricity excitation of giant planets. 

In 3D simulations the planet remains on a quasi-circular orbit, probably because there is still enough gas in the gap to damp continuously its eccentricity. The meridional circulation may also play a role in unsaturating corotation resonances. The vortex tends to migrate inwards by exerting a torque on the outer disc and therefore can accompany the planet in its inward migration. Thus, initially planet migration is inward and occurs at the speed of vortex migration, which is unrelated to the viscosity of the disc \citep{Paardekooperetal10}. We call this {\it vortex-driven migration} (see Sect.~\ref{Def}). However, because the disc is inviscid, the torque exerted by the vortex opens a secondary gap just beyond the vortex's orbit. The vortex then spreads radially and becomes less pronounced. These processes partially deplete the disc just beyond the planet's gap and, consequently, the planet starts to feel a positive torque from the inner disc that exceeds the negative torque from the outer disc: its migration is reversed. By being pushed out, eventually the planet adjusts its position relative to the inner and outer discs to balance the two torques and its migration stops. If the mass of the disc is reduced, the relative strength of the vortex is also reduced. Consequently, no significant secondary gap is opened; migration is not reversed, but it slows down to a rate smaller than 1 au/My; it should eventually stop on the long-term. 

In both cases, these results can potentially explain the lack of large-scale migration of giant planets. If inward migration is short-ranged and eventually slows down, or reverses and stops, the presence of numerous giant planets beyond 1~au from the central star (i.e. cold-Jupiters) becomes consistent with the idea that the sweet-spot for giant planet formation is the disc's snowline, which typically is in the same region or only slightly beyond. This result is very promising to solve the giant planet migration problem, but it needs to be confirmed using more realistic discs. In fact, the disc we have adopted { with very low viscosity and therefore negligible radial transport} would be incompatible with the mass accretion rates that are typically observed for young stars. If the protoplanetary discs have low viscosities because of the lack of magneto-rotational instability, most likely the accretion of gas towards the star occurs due to laminar stresses where angular momentum is removed in disc winds (see \citet{Turner2014} for a review). In a future work we will study giant planet migration in these  discs. Nevertheless, the results presented in this paper open a new perspective and will serve as a basis for comparison. 

In the final part of the paper we have also stressed that the result of slow, or reversed, giant-planet migration for low viscosities holds only in the inner part of the disc. The outer part of the disc is closer to the gravitationally instability limit and therefore can respond very differently to the presence of a giant planet. Unfortunately, our 3D code cannot account for self-gravity in its present form. A future work will be devoted to performing 3D self-gravitating simulations. Here, by using 2D self-gravitating simulations we observed that the disc may develop a strong $m=1$ eccentric mode, which in turn excites the eccentricity of the planet, triggering fast inward migration. This result suggests that giant planet migration may be fast in the outer disc and slow in the inner disc, so that giant planets, wherever they form, should all pile-up in the cold-Jupiter region. If this is true, we would then predict a strong drop-off in the giant planet radial distribution, with only the most massive planets --the least sensitive to planet-disc interactions-- remaining at large distances where they can be probed by direct imaging surveys.  The stakes are high, so it is important to confirm or improve on these preliminary conclusions with 3D self-gravitating simulations. 

\begin{acknowledgements}
 We acknowledge support by DFG-ANR supported GEPARD project
 (ANR-18-CE92-0044 DFG: KL 650/31-1). We also acknowledge HPC resources from GENCI DARI n.A0080407233 and from "Mesocentre SIGAMM" hosted by Observatoire de la C\^ote d'Azur. LE whish to thank Alain Miniussi for maintainance and re-factorization of the code FARGOCA. RPN acknowledges support from the STFC through the Consolidated Grants ST/M001202/1 and ST/P000592/. This research utilised Queen Mary's Apocrita HPC facility, supported by QMUL Research-IT. TR acknowledges support by the DFG research  group FOR2634: "Planet Formation Witnesses and Probes: Transition Disks" under grant KL650/29-1
\end{acknowledgements}

\begin{appendix}
\section{Sensitivity of results to the cooling time}
\label{sec:Appendix1}
\subsection{2D simulations}
\label{sec:Appendix2D}

In addition to running simulations for cooling time $\tau_{\rm c}=1$~orbit, we also ran simulation suites for self-gravitating and non-self-gravitating models with $\tau_{\rm c}=0.01$ and $\tau_{\rm c}=10$~orbits. We considered the disc masses of the $M$, $M/2$ and $M/4$ simulations. Our results indicate that for 2D simulations, the detailed outcome depends on the cooling time in a rather complicated manner.

\subsubsection{$\tau_{\rm c}=0.01$ orbits}
\begin{figure}
\centering
       \includegraphics[width=\hsize]{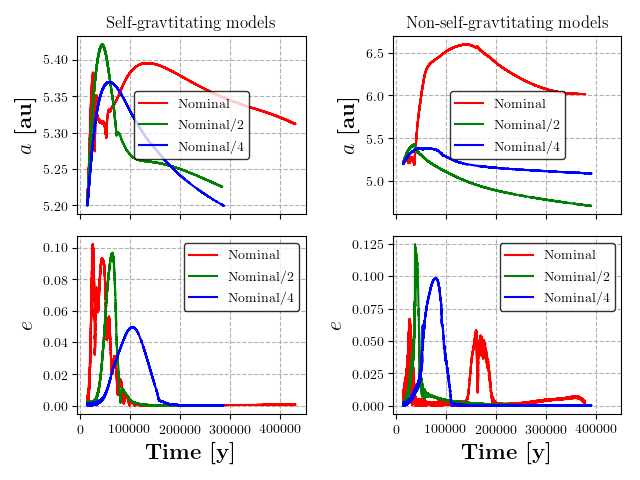}
          \caption{Left panels: Evolution of semi-major axes and eccentricities for planets in simulations $M_{\rm 2D}$,  $M_{\rm 2D}/2$  and $M_{\rm 2D}/4$ with cooling time $\tau_{\rm c}=0.01$~orbit. Right panels: Same as left panels except for the corresponding non-self-gravitating disc models. }
         \label{Fig:aemigr2Dtcool0.0628}
   \end{figure}

\begin{figure*}
\centering
    \includegraphics[width=0.95\textwidth]{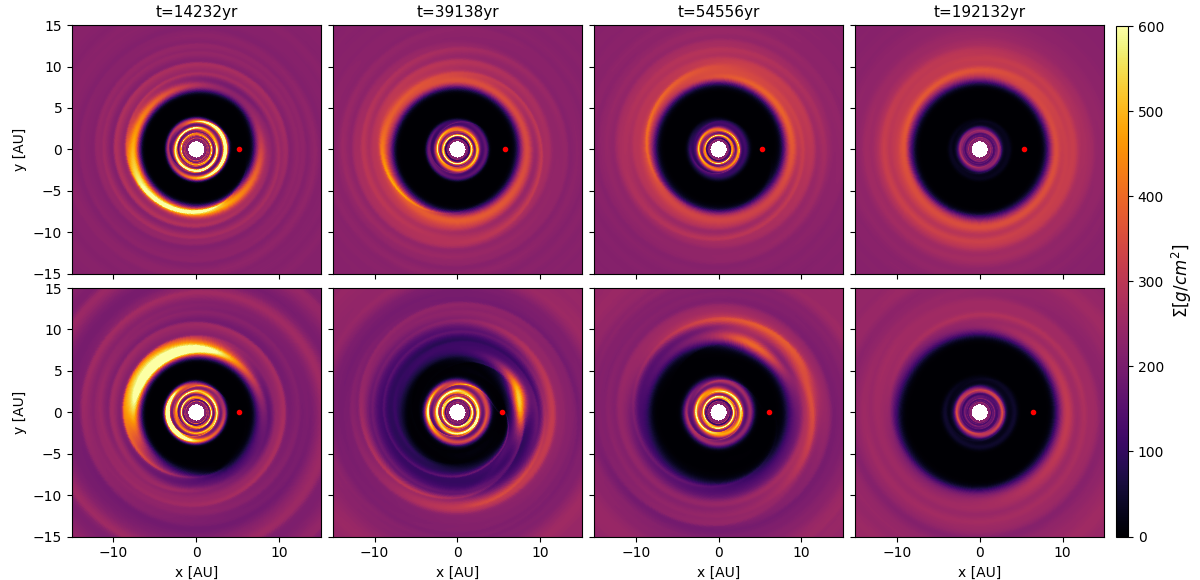}
   \caption{Contours of surface density for simulation $M_{\rm 2D}$ with $\tau_{\rm c}=0.01$ orbits and a migrating Jupiter-mass planet. The panels correspond to different times, reported on top of each panel. Top panels: Self-gravitating discs. Bottom panels: Non-self-gravitating discs.}
   \label{Fig:Sigma2D_appendix1}
\end{figure*}

Fig.\ref{Fig:Sigma2D_appendix1} shows the surface density evolution for the $M_{\rm 2D}$ simulation with cooling time of 0.01~orbits. The top panels are for the self-gravitating run and the bottom panels are for the non-self-gravitating disc. This figure should be compared with Fig.\ref{fig:surfdenPlanet5.2AU} in the main text, which shows the results for $\tau_{\rm c}=1$~orbit. The shorter cooling time gives rise to different qualitative behaviour of the disc and the embedded planet, which is released from a fixed orbit at the time corresponding to the left-most panels in Fig.~\ref{Fig:Sigma2D_appendix1}. The self-gravitating run shows the formation of a strong vortex at the outer edge of the gap, and this remains relatively narrow in its radial extent and stays at the gap edge without migrating. Over time the vortex weakens and dissipates, and appears to have almost completely dispersed in the right-most panel of Fig.~\ref{Fig:Sigma2D_appendix1}. Analysis of the disc at late times, extending beyond the time of the right-most panel of Fig.~\ref{Fig:Sigma2D_appendix1} to $\sim 500,000$~yr, indicates that a vortex remains at the pressure bump at the outer edge, but with a strength that waxes and wanes intermittently throughout the run. The non-self-gravitating run results in the formation of a more radially extended vortex, and the evolution here is much more similar to the 3D results described in section~\ref{sec:3DNominal}. The radial width of the vortex is large enough for the excitation of spiral waves to drive its migration into the gap, while at the same time the vortex creates a secondary gap (lower panels, second from left). The third panel shows that the formation of a secondary gap also leads to the formation of a secondary pressure bump, and this forms a second vortex via the Rossby wave instability. Eventually, the primary vortex disperses, leaving the weaker secondary vortex orbiting at the gap edge (fourth panel). This vortex persists for the whole simulation. Hence, we see that in these self-gravitating and non-self-gravitating runs, the initial evolution of the system is quite different because of the formation of a large, migrating vortex in the latter simulation, but the long term behaviour is actually very similar.

We do not show figures for the $M_{\rm 2D}/2$ and $M_{\rm 2D}/4$ runs, but instead comment that the qualitative differences between self-gravitating and non-self-gravitating models are much smaller than shown in Fig.~\ref{Fig:Sigma2D_appendix1}. Reducing the disc mass reduces the importance of the indirect term, and hence the radial width of the vortex that forms is smaller in these cases. Hence, the outer gap-edge vortex does not migrate quickly into the gap, but remains at the gap edge and slowly dissipates while generating a shallow secondary gap. The main difference between the self-gravitating and non-self-gravitating models is that the vortex weakens more quickly in the self-gravitating models. The vortex in the non-self-gravitating models is persistent until the ends of the simulations, providing a source of weak effective viscosity via the generation of Reynolds stresses, whereas the vortex in the self-gravitating discs appears to be more intermittent in its behaviour.

Fig.~\ref{Fig:aemigr2Dtcool0.0628} shows the evolution of semi-major axes and eccentricities for the $M_{\rm 2D}$, $M_{\rm 2D}/2$ and $M_{\rm 2D}/4$  runs with $\tau_{\rm c}=0.01$. The self-gravitating models show the planet initially adjusts its orbital location in the gap due to an imbalance of torques, and experiences a phase of eccentricity growth due to direct interaction with the vortex. The eccentricity, however, is eventually damped by interaction with the disc as the vortex weakens, and remains at a very low value ($\sim 5 \times 10^{-3}$) for the remainder of the run. During this low eccentricity phase, we see that the migration rate is very slow, corresponding to migration over a distance of $1$~au in $4 \times 10^6$~yr (i.e. a migration speed of 0.25 au per Myr) for the nominal model. The orbital evolution during the $M_{\rm 2D}$ simulation is initially more extreme because of the strong interaction with the migrating vortex, whereas the evolution in the $M_{\rm 2D}/2$ and $M_{\rm 2D}/4$ simulations are similar to the self-gravitating models. Over the long term, however, we see that in these models the planet also exhibits low eccentricities and very slow migration rates. Hence, the $\tau_{\rm c}=0.01$ self-gravitating and non-self-gravitating models during long-term evolution exhibit the low eccentricity mode of migration observed in the $M_{\rm 3D}$ simulations described in section~\ref{sec:3DNominal}. This maintenance of low eccentricity in an inviscid disc is not expected, since the corotation resonances that are responsible for damping eccentricity are expected to saturate and switch off. The persistence of vortices at the outer gap edge, however, seems to provide a source of effective viscosity that allows these resonances to actively damp the eccentricity.

\subsubsection{$\tau_{\rm c}=10$ orbits}

\begin{figure*}
\centering
    \includegraphics[width=0.95\textwidth]{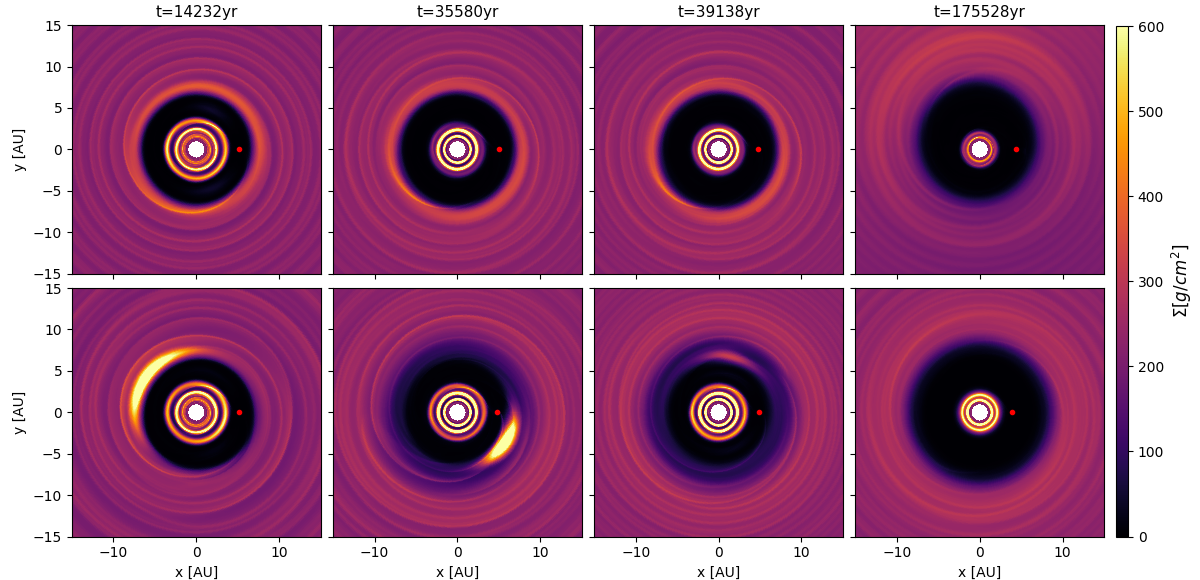}
   \caption{Contours of surface density for simulations $M_{\rm 2D}$   with $\tau_{\rm c}=10$ orbits and a migrating Jupiter-mass planet. The panels correspond to different times, reported on top of each panel. Top panels: Self-gravitating discs. Bottom panels: Non-self-gravitating discs.}
   \label{Fig:Sigma2D_appendix2}
\end{figure*}

Fig.~\ref{Fig:Sigma2D_appendix2} shows the surface density evolution for the $M_{\rm 2D}$sg (top panels) and $M_{\rm 2D}$ models (bottom panels). The self-gravitating disc shows similar behaviour to that seen previously for both $\tau_{\rm c}=0.01$ and $\tau_{\rm c}=1$, namely the formation of a radially-narrow vortex that remains at the gap edge while weakening and eventually dissipating due to the onset of the Rayleigh instability. For $\tau_{\rm c}=1$ and 10 the vortex dissipates much more quickly than in the $\tau_{\rm c}=0.01$ run. The non-self-gravitating run shows the formation of a radially-extended vortex which forms a significant secondary gap and migrates into the planet-induced gap before dispersing. This occurs relatively quickly in the simulation, and leaves behind a smooth outer disc without any evidence for additional vortices being present (unlike in the $\tau_{\rm c}=0.01$ run). 
\begin{figure}
\centering
       \includegraphics[width=\hsize]{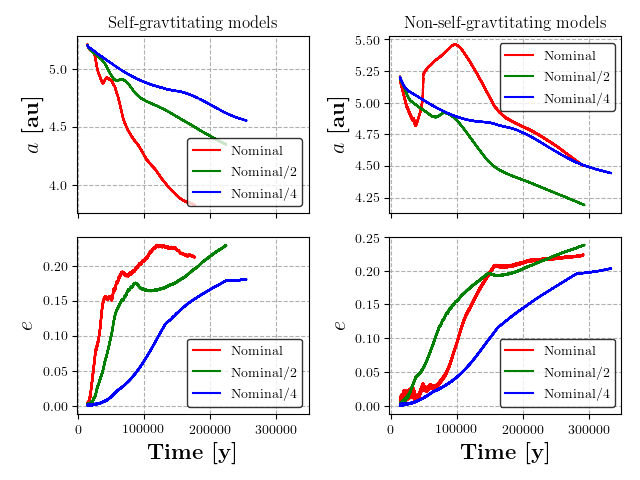}
          \caption{Left panels: Evolution of semi-major axes and eccentricities for planets in self-gravitating disc simulations $M_{\rm 2D}$, $M_{\rm 2D}/2$ and $M_{\rm 2D}/4$ where $\tau_{\rm c}=10$~orbit. Right panels: Same as left panels except for the corresponding non-self-gravitating disc models. }
         \label{Fig:aemigr2Dtcool62.8}
   \end{figure}
The evolution of the discs for the $M_{\rm 2D}/2$ and $M_{\rm 2D}/4$ models show much stronger convergence with respect to whether self-gravity is included or not. Here the vortex at the outer edge is radially narrow and does not migrate into the gap and strongly interacts with the planet. Instead it sits at the gap edge in all cases, weakening with time and eventually dissipating.

Fig.~\ref{Fig:aemigr2Dtcool62.8} shows the evolution of semi-major axes and eccentricities for all runs with $\tau_{\rm c}=10$. Apart from the initial phase of evolution of the $M_{\rm 2D}$sg case, where the planet interacts strongly with the vortex, we see that these models all display the high-eccentricity mode of migration described in the main text for the models with $\tau_{\rm c}=1$. The mean migration rates vary somewhat between the runs, but correspond to migration over a distance of 1~au in $150,000$--400,000~yr. Hence, Jovian mass planets in these models can migrate to the star within typical disc lifetimes.

\subsection{3D simulations}
\label{sec:Appendix3D}

\begin{figure*}
\centering
    \includegraphics[width=0.95\textwidth]{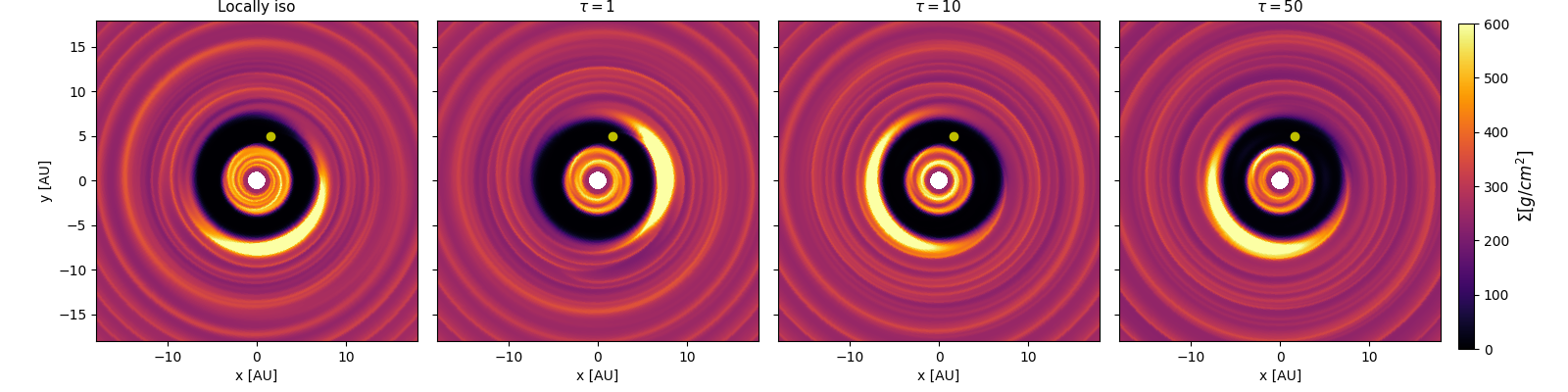}
   \caption{Contours of surface density for the $M_{\rm 3D}$ simulations with the locally isothermal equation of state and for increasing values of $\tau_ {\rm c}$  (in orbits). All the panels correspond to $t=8296 yr$ (the same time as for the middle panels of Fig.\ref{Fig:NominalSG-NOSG-r=1}, for comparison).}
   \label{Fig:Sigma3D_appendix2}
\end{figure*}
In addition to the Nominal 3D case with $\tau_{\rm c}=1$ we ran simulations with fixed planet only for the locally isothermal equation of state and for $\tau_{\rm c}=10$ and $\tau_{\rm c}=50$ orbits.

We can consider the locally isothermal case equivalent to having an extremely short cooling time, so that in Fig.\ref{Fig:Sigma3D_appendix2} we show results (from left to right panels) from short cooling times to high values such as $\tau _{\rm c}=50$ orbits. 
\par In all cases a big vortex forms at the outer edge of the gap and qualitatively all the panels are very similar. It appears that 3D models are much less sensitive to cooling times than 2D ones.
{We remark that for large values of the cooling time, the vertical temperature profile is not damped efficiently toward the initial condition:
 $h_0=0.05$. Actually, for $\tau _{\rm c}=50$ orbital periods the aspect ratio fluctuates around a mean value of $0.057$ at the time snapshot of Fig.\ref{Fig:Sigma3D_appendix2}. Qualitatively there is no influence on
 vortex formation but a more detailed study of the role of the thermal structure on vortex formation and planet migration goes beyond the scope of the present paper.}
\par Due to the expensive computation required for 3D migration, we decided to limit our analysis of the sensitivity to the cooling time to fixed planets. However, the similarity in vortex formation indicates that the migration histories would probably be very similar to the
case $\tau_{\rm c}=1$ orbit described in Section \ref{sec:migrr1}.

\section{Numerics}
\label{resolution}
We have seen above that vortex-driven migration occurs if gaps are only moderately deep, so  it is important to check  the dependence of results with respect to the choice of the smoothing of the planetary potential. 
Moreover, vortex formation and evolution may depend
on numerical resolution. It is therefore important to check also the validity of results with respect to the grid stepsize. We present these tests below.

\par In all the simulations we smooth the potential of the planets for disc elements with distance $d$ from the planet: $d<\epsilon$. We have considered  $\epsilon = 0.8 R_H$ in our nominal $M_{\rm 3D}$ simulations. 
In order to check the sensitivity to the smoothing length we have run a simulation at nominal resolution  with  $\epsilon = 0.4 R_H$ for
800 orbits.

\begin{figure}
\centering
   \includegraphics[width=\hsize]{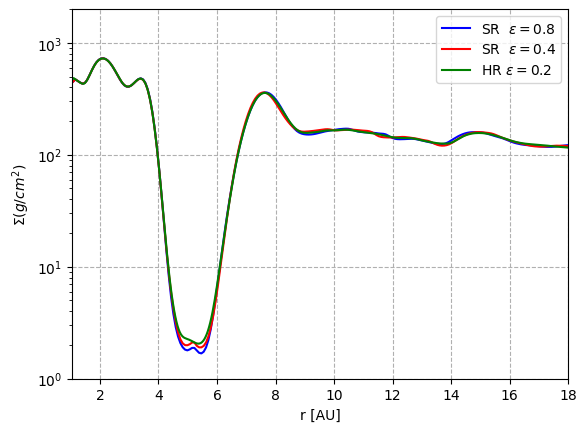}
      \caption{Surface density profiles { at $t=10000$ years} for standard resolution { non viscous} 3D simulations with different smoothing length and for an additional run with double resolution and smoothing length $\epsilon=0.2$ Hill radii. We call  "SR"  the standard resolution simulations and we use the label "HR" for the high resolution one. { The "HR" simulation is a  restart of the "SR" case with $\epsilon = 0.4 R_H$ (red curve) at $t=10^4$ years.  The green curve corresponds to  100 orbits after the restart}.  }
         \label{Fig:Smoothing} 
\end{figure}

Fig.\ref{Fig:Smoothing} shows that the surface density profiles 
for the nominal simulation (blue curve) and the new one with
reduced smoothing length (red curve) at the time of planet release.
As one can see, the two curves are essentially identical.
Thus, we can conclude that the limited depth of the gap is not due to
an excessive smoothing but corresponds to the real dynamic of the gas.

\begin{figure}
\centering
   \includegraphics[width=\hsize]{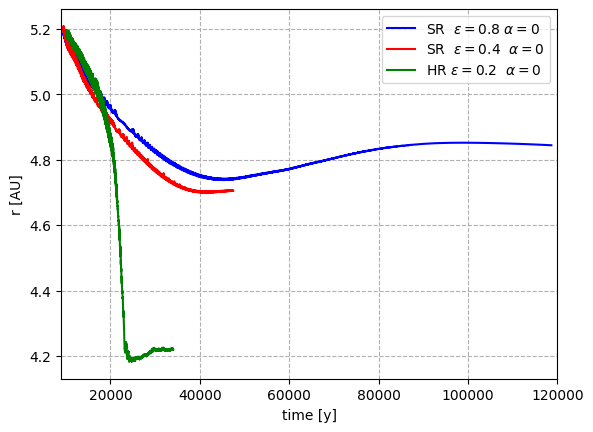}
      \caption{Evolution of the planet's semi-major axis for the SR $M_{\rm 3D}$ simulation with different smoothing lengths and for the additional HR simulation with smoothing length
      $\epsilon=0.2$ Hill radii.}
         \label{Fig:App3Dmigrationalpha0} 
\end{figure}

When the planet is released (Fig.\ref{Fig:App3Dmigrationalpha0}, red curve)
the migration is very similar, although slightly faster than the nominal simulation. Importantly, we see the same reversal of migration direction, although this happens slightly earlier with the planet being slightly closer to the Sun.

\begin{figure}
\centering
   \includegraphics[width=\hsize]{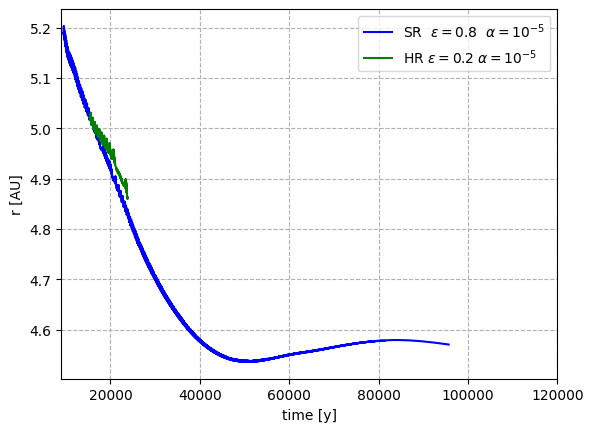}
      \caption{Evolution of the planet's semi-major axis for the SR $M_{\rm 3D}$ simulation with viscosity $\alpha = 10^{-5}$ and for the additional HR simulation with smoothing length
      $\epsilon=0.2$ Hill radii.}
         \label{Fig:App3Dmigrationalpha5} 
\end{figure}

Testing the effect of resolution is hard because of the long integration times. Thus we have  restarted the simulation with $\epsilon = 0.4 R_H$   with double resolution  at $t=10^4$ years when the planet is still on a fixed orbit.
We recall that the nominal $M_{\rm 3D}$ simulation had a resolution $(N_r,N_{\theta},N_{\varphi})=(568,16,360)$. In the high resolution simulations we reduced further the smoothing length  to $\epsilon = 0.2 R_H$.
The surface density profile obtained after 100 orbits is shown by the green curve in Fig.\ref{Fig:Smoothing} and is extremely similar to those of the two other simulations.
We then released the planet.
{On a timescale of about 800 orbits the migration rate is indistinguishable from that of the previous simulations
(Fig.\ref{Fig:App3Dmigrationalpha0}).
Then, at time $\sim 19000$ y  we observe an  episode of rapid inward migration (and eccentricity increase up to 0.07),  before turning to slow outward migration as in the nominal resolution case.
The outward migration phase starts at $t=\sim 23000$ years when the vortex is weakened by radial spreading and the eccentricity is damped to $\sim 0.001$.  Finally, planet's migration stops at a position in which the  torques from the inner and outer discs  balance.
From this test we conclude that in the non viscous  case the dynamics described in the main paper is not an artefact of resolution, although the evolution of the planet can change at the quantitative level using different resolutions.
To obtain a quantitative convergence  of results with respect to resolution and smoothing length a low prescribed viscosity may be required. \par
To do this we have also run the double resolution test by considering the low viscosity $\alpha =10^{-5}$ simulation $M_{{\rm 3D}\alpha 5}$ for which we have observed vortex-driven migration (see fig.\ref{Fig:speed3D5.2auAlpha}).
 We have restarted the simulation with $\alpha=10^{-5}$  by doubling the resolution at $t \sim 16000$ years when the planet is already in the migration phase. 
This choice is motivated on the one hand by  computational cost (about 1 orbit at 5.2 au per hour using 240 cores with Hybrid  OpenMP+MPI parallelization), on the other hand by the results obtained with double resolution in the inviscid case.
In fact, as stated above, up to $t\sim 19000$ years migration doesn't show any dependency on resolution (Fig.\ref{Fig:App3Dmigrationalpha0}).\par
Fig.\ref{Fig:App3Dmigrationalpha5} shows that the migration rate is not affected by the change in the resolution (we did not observe any increase in eccentricity either). We have integrated this case up to
$t \sim 23000$ years. We consider this integration time  
(corresponding to the end of the fast inward migration phase observed for the non viscous case) long enough to
prove quantitative robustness of results with respect to
resolution and smoothing length.\par
We conclude that a small prescribed  viscosity ($\alpha=10^{-5}$) is necessary to show convergence of results with numerical resolution also at the  quantitative level, confirming the results of the paper.}

\end{appendix}


\end{document}